\documentclass[twocolumn]{aastex631}
\usepackage{amsmath}
\usepackage{CJK}
\usepackage{xcolor}

\begin{document}
\begin{CJK*}{UTF8}{gbsn}

\title{On the evolution of the Anisotropic Scaling of Magnetohydrodynamic Turbulence in the Inner Heliosphere.}

\author[0000-0002-1128-9685]{Nikos Sioulas}
\affiliation{Department of Earth, Planetary, and Space Sciences, University of California, Los Angeles \\
Los Angeles, CA 90095, USA}

\author[0000-0002-2381-3106]{Marco Velli}
\affiliation{Department of Earth, Planetary, and Space Sciences, University of California, Los Angeles \\
Los Angeles, CA 90095, USA}
\affil{International Space Science Institute, 3012 Bern
Switzerland}

\author[0000-0001-9570-5975]{Zesen Huang (黄泽森)}
\affiliation{Department of Earth, Planetary, and Space Sciences, University of California, Los Angeles \\
Los Angeles, CA 90095, USA}

\author[0000-0002-2582-7085]{Chen Shi (时辰)}
\affiliation{Department of Earth, Planetary, and Space Sciences, University of California, Los Angeles \\
Los Angeles, CA 90095, USA}


\author[0000-0002-4625-3332]{ Trevor A. Bowen}
\affiliation{Space Sciences Laboratory, University of California, \\ Berkeley, CA 94720-7450, USA}

\author[0000-0003-4177-3328]{B. D. G. Chandran}
\affiliation{Space Science Center, University of New Hampshire, Durham, NH 03824, USA}

\author[0000-0002-8921-3760]{Ioannis Liodis}
\affiliation{Deprtment of Physics,  Aristotle University of Thessaloniki\\
GR-52124 Thessaloniki, Greece }

\author[0000-0001-7222-3869]{Nooshin Davis}
\affiliation{Space Science Center and Department of Physics, University of New Hampshire, Durham, NH 03824, USA}

\author[0000-0002-1989-3596]{Stuart D. Bale}
\affil{Physics Department, University of California, Berkeley, CA 94720-7300, USA}
\affil{Space Sciences Laboratory, University of California, Berkeley, CA 94720-7450, USA}

\author[0000-0002-7572-4690]{T. S. Horbury}
\affiliation{The Blackett Laboratory, Imperial College London, London, UK}

\author[0000-0002-4401-0943]{Thierry {Dudok de Wit}}
\affil{LPC2E, CNRS and University of Orl\'eans, Orl\'eans, France}

\author[0000-0001-5030-6030]{Davin Larson}
\affiliation{Space Sciences Laboratory, University of California, Berkeley, CA 94720-7450, USA}

\author[0000-0002-7728-0085]{Michael L. Stevens}
\affiliation{Harvard-Smithsonian Center for Astrophysics, \\ Cambridge, MA 02138, USA}

\author[0000-0002-7077-930X]{Justin Kasper}
\affiliation{BWX Technologies, Inc., Washington DC 20002, USA}
\affiliation{Climate and Space Sciences and Engineering, University of Michigan, Ann Arbor, MI 48109, USA}

\author[0000-0002-5982-4667]{Christopher J. Owen}
\affiliation{Mullard Space Science Laboratory, University College London, Dorking, RH5 6NT, UK}

\author[0000-0002-3520-4041]{Anthony Case}
\affiliation{Smithsonian Astrophysical Observatory, \\
Cambridge, MA 02138, US}

\author[0000-0002-1573-7457]{Marc Pulupa}
\affil{Space Sciences Laboratory, University of California, Berkeley, CA 94720-7450, USA}

\author[0000-0003-1191-1558]{David M. Malaspina}
\affil{Astrophysical and Planetary Sciences Department, University of Colorado, \\ Boulder, CO, USA}
\affil{Laboratory for Atmospheric and Space Physics, University of Colorado, \\ Boulder, CO, USA}

\author[0000-0002-0396-0547]{Roberto Livi}
\affiliation{Space Sciences Laboratory, University of California, Berkeley, CA 94720-7450, USA}

\author[0000-0003-0420-3633]{Keith Goetz}
\affiliation{School of Physics and Astronomy, University of Minnesota, Minneapolis, MN 55455, USA}

\author[0000-0002-6938-0166]{Peter R. Harvey}
\affil{Space Sciences Laboratory, University of California, Berkeley, CA 94720-7450, USA}

\author[0000-0003-3112-4201]{Robert J. MacDowall}
\affil{Solar System Exploration Division, NASA/Goddard Space Flight Center, Greenbelt, MD, 20771}


\begin{abstract}
We analyze a merged  Parker Solar Probe ($PSP$) and  Solar Orbiter ($SO$) dataset covering heliocentric distances $13 \ R_{\odot} \lesssim R \lesssim 220$ $R_{\odot}$ to investigate the radial evolution of power and spectral-index anisotropy in the wavevector space of solar wind turbulence. Our results show that anisotropic signatures of turbulence display a distinct radial evolution when fast, $V_{sw} \geq ~ 400 ~km ~s^{-1}$, and slow, $V_{sw} \leq ~ 400 ~km ~s^{-1}$, wind streams are considered. The anisotropic properties of slow wind in Earth orbit are consistent with a ``critically balanced'' cascade, but both spectral-index anisotropy and power anisotropy diminish with decreasing heliographic distance. Fast streams are observed to roughly retain their near-Sun anisotropic properties, with the observed spectral index and power anisotropies being more consistent with a ``dynamically aligned'' type of cascade, though the lack of extended fast-wind intervals makes it difficult to accurately measure the anisotropic scaling. A high-resolution analysis during the first perihelion of PSP confirms the presence of two sub-ranges within the inertial range, which may be associated with the transition from weak to strong turbulence. The transition occurs at $\kappa d_{i} \approx 6 \times 10^{-2}$, and signifies a shift from -5/3 to -2 and -3/2 to -1.57 scaling in parallel and perpendicular spectra, respectively. Our results provide strong observational constraints for anisotropic theories of MHD turbulence in the solar wind.

\end{abstract}


\correspondingauthor{Nikos Sioulas}
\email{nsioulas@ucla.edu}

\section{Introduction} \label{sec:intro}

Magnetohydrodynamic (MHD) turbulence  is relevant to a wide range of astrophysical systems such as stellar coronae, stellar winds, and the interstellar medium. A large-scale magnetic field $\boldsymbol{B}_{0}$  \footnote{Magnetic fields are presented in Alfv\'en velocity units, e.g., $\boldsymbol{b} \rightarrow \boldsymbol{b}/\sqrt{4 \pi \rho}$, $\boldsymbol{B}_{0} \equiv \boldsymbol{V}_{a}$ } is often present \citep{parker_magnetic_fields,biskamp_magnetohydrodynamic_2003} and the fluctuations are typically observed to be mostly incompressible. The incompressible MHD equations are better expressed using Elsasser variables, $\boldsymbol{z}^{\pm} = \boldsymbol{v} \pm \boldsymbol{b}$ \citep{elsasser_1950}, where the nonlinear term for each variable may be written
$\partial_{t}\boldsymbol{z}^{\pm} \sim -  \boldsymbol{z}^{\mp} \cdot \nabla \boldsymbol{z}^{\pm}$ (here
$\sim$ means proportional up to a projection operator ensuring incompressibility): nonlinear effects therefore require interactions between fluctuations with opposite signs of cross-helicity. Based on the weak interaction of oppositely moving Alfv\'enic wavepackets in a strong background magnetic field, $\delta v, \delta b \ll B_{0}$, i.e., 
assuming that the wave propagation, $\tau_{A}(\boldsymbol{\kappa}) = 1/|\boldsymbol{B} \cdot \boldsymbol{k}|$
is shorter than the nonlinear decay time $\tau_{nl}(\boldsymbol{\kappa}) \approx 1/(k \cdot \delta u_{k})$, where $\delta u_{k}$ is
the average velocity fluctuation at scales $\ell \sim 1/|\boldsymbol{k}|$, the turbulent cascade will be slowed relative to hydrodynamic turbulence \citep{iroshnikov_turbulence_1963,kraichnan_inertial-range_1965}. Assuming homogeneity, isotropy, $\boldsymbol{B} \cdot \boldsymbol{k} \rightarrow  B \cdot k$, and scale locality of the interactions, simple dimensional analysis then leads to the prediction of the inertial range omnidirectional power-spectrum, $E(k) \propto k^{-3/2}$. Magnetic fields, however, cannot be eliminated via Galilean transformations of MHD equations, as opposed to mean velocity fields, $\boldsymbol{V}_{0}$, resulting in strongly anisotropic turbulent dynamics \citep{Montgomery_turner_1981} \citep[see also reviews by][and references therein]{Schekochihin_2009_review,  Oughton_anisotropy_review}. In particular, conservation of energy and momentum during wave-wave interactions (more specifically, a wave - 2D perturbation interaction, \citep[see,][]{1995ApJ...447..706M}) allows power to cascade down to smaller scales perpendicular to $\boldsymbol{B}_{0}$, resulting in a two-dimensionalization of the turbulence spectrum in a plane transverse to the locally dominant magnetic field while at the same time inhibiting spectral energy transfer along the direction parallel to the field. \citep{shebalin_matthaeus_montgomery_1983, Ng_Bhattacharjee, Galtier_2000_anisotropic}. 

A multitude of observational and numerical studies have  investigated the manifestations of anisotropy in the presence of an energetically significant mean magnetic field e.g., anisotropy in correlation functions, power at a fixed scale, spectral indices, intermittency \citep{belcher_large-amplitude_1971, Matthaeus_1990_anisotropy, Bieber_anisotropy, Maron_2001, Weygand_anisotropy, Beresnyak_2010, osman_intermittency_2012, wicks_alignmene_2013, Chandran_Perez_2019, Pine_2020,Bandyopadhyay_2021, Zank_2022,  Sioulas_2022_intermittency,  Chhiber_2022, Dong_22_largest_mhd_turb}. A comprehensive overview of the various forms of anisotropy can be found in \cite{Horbury_anisotropy}. \par

Using in-situ observations in the solar wind \citep{horbury_anisotropic_2008, Podesta_2009, Chen_2010_anisotropy_multispacecraft} explored the dependence of the scaling index of the magnetic power spectrum's inertial range, $\alpha_{B}$, on the field/flow angle $\theta_{BV}$. An essential nuance in observing scale-dependent anisotropy involves the necessity of measuring parallel correlations along a local, scale-dependent mean magnetic field, $\boldsymbol{B}_{\ell}$, instead of the global mean magnetic field, as emphasized by \citep{Cho_Vishniac_2000, Gerick_2017}, \citep[see also, review by][ and references therein]{Schekochihin_2022}. The aforementioned studies suggest inertial range spectral indices of $-2$ and $-5/3$ for flow directions parallel ($\Theta_{BV} \approx 0^{\circ}$) and perpendicular ($\Theta_{BV} \approx 90^{\circ}$) to the mean magnetic field, respectively.  These observations were interpreted as supporting evidence for the $\it{critical~ balance}$ (CB) theory \citep{1994_goldreich, goldreich_toward_1995, goldreich_magnetohydrodynamic_1997} \footnote{Heavily influenced by the work of \cite{higdon_anisotropic}}, which is based on the conjecture that the inertial range dynamics of MHD turbulence with vanishing cross-helicity ($\sigma_{c} \approx 0$), later extended to imbalanced cascades \citep{Lithwick_2007_imbalanced_critical_balance}, are governed by wavevector modes for which rough equality between $\tau_{A}(\boldsymbol{\kappa})$ and $\tau_{nl}(\boldsymbol{\kappa})$, $\tau_{A}(\boldsymbol{k}) \approx \tau_{nl}(\boldsymbol{|k|})$ holds. As a result, the relationship between the parallel and perpendicular wavevectors follows an anisotropic scaling, $\kappa_{||} \sim \kappa_{\perp}^{2/3}$. Based on this scaling, we expect the magnetic fluctuation spectra to follow scalings of: $E(k_{\perp}) \propto k_{\perp}^{-5/3}$ and $E(k_{||}) \propto k_{||}^{-2}$. However, the $\it{dynamic~ alignment}$ conjecture \citep{boldyrev_2006, masson_2006, Perez_Boldyrev_Extend} suggests that, as the energy cascades to smaller scales, velocity and magnetic field fluctuations in the plane perpendicular to $\boldsymbol{B}_{\ell}$ will align to within a smaller angle $\phi$, resulting in weaker nonlinear interactions and a flatter perpendicular inertial range spectrum, $E(k_{\perp}) \propto k_{\perp}^{-3/2}$. In contrast, the field parallel spectrum remains unchanged, $E(k_{||}) \propto k_{||}^{-2}$. 
Other models of turbulence, such as the 2D plus slab model \cite{Zank_2020_transition} lead to perpendicular and parallel spectra that can range between $5/3$ and $3/2$ in the perpendicular direction and $5/3$ and $2$ in the parallel direction.  

\par

Recent observations from the Parker Solar Probe ($PSP$) and Solar Orbiter ($SO$) missions have provided the opportunity to investigate the radial evolution of turbulence in the inner heliosphere. It was shown that the inertial range of the magnetic spectrum grows with distance, progressively extending to larger spatial scales \citep{sioulas_turb_22_no1} while at the same time steepening from a scaling of $\alpha_{B} = -3/2$ at approximately 0.06 au towards the Kolmogorov scaling of $\alpha_{B} =-5/3$ \citep{chen_evolution_2020, alberti_scaling_2020, telloni_evolution_2021, shi_alfvenic_2021, Zhao_2022ApJ}. The rate at which the spectrum steepens has also been found to be related to the Alfv\'enic content and magnetic energy excess of the fluctuations \citep{sioulas_turb_22_no1}. On the contrary, the spectral index of the velocity spectrum in the inertial range  has consistently been found to be close to $\alpha_{v} = -3/2$, regardless of the distance from the Sun \citep{shi_alfvenic_2021}.


In this study, we aim to understand the radial evolution of anisotropic magnetic turbulence in the inner heliosphere. To do this, we analyze data from the PSP and SO missions covering heliocentric distances $13 \ R_{\odot}\lesssim R\lesssim 220$~$R_{\odot}$ using wavelet analysis. This technique allows us to decompose the magnetic field timeseries into scale-dependent background and fluctuations, and study the dependence of the turbulence properties on the field/flow angle $\theta_{BV}$.\par

The rest of the paper is structured as follows: In Section \ref{sec:wavelet_analysis}, we provide background on wavelet analysis and introduce the anisotropy diagnostics used in this study. Section \ref{sec:Data} presents the selection and processing of the data. The results of this study are presented in Section \ref{sec:results}, with a focus on high-resolution data obtained during the first perihelion of PSP in Subsection \ref{subsec:E1}, and the radial evolution of magnetic field anisotropy investigated in Subsection \ref{subsec:radial_evolution_of_turbulence}. In Section \ref{sec:Comparison}, we compare our findings to previous relevant studies in order to advance our understanding of the topic and validate our conclusions. The discussion of the results and conclusions are provided in Sections \ref{sec:discussion} and \ref{sec:Conclusions}, respectively.








\section{Diagnostics}\label{sec:wavelet_analysis}
 
\subsection{Wavelet analysis  $\&$ estimation of Power Spectral Density (PSD)}\label{subsec:wavelet_analysis}

Anisotropy in turbulence represents a local property that relies on both the position and scale. The turbulent fluctuations at a given scale $\ell$ are greatly influenced by the local mean magnetic field of a size that ranges between $3-5\cdot\ell$ \citep{Gerick_2017}. To analyze anisotropy, wavelet analysis has proven to be a useful technique as it allows signal decomposition into components that are localized both in time and wavelet scale. Recently, the continuous wavelet transform (CWT) has been extensively utilized to estimate the power of magnetic field fluctuations as a function of the direction of the local mean magnetic field \citep{Podesta_2009, wicks_anisotropy1}. For a discrete set of measurements such as the time series of the $i$-th component of the magnetic field $B_i$, where $i=R, ~T, ~N$ and resolution $\delta t$ , the wavelet transform is defined as:

\begin{equation}
    \omega_{i}(\ell, ~ t_{n}) = \sum_{j=0}^{N-1}B_{i}(t_{j})\psi^{\ast}(\frac{t_{j} -t_{n}}{\ell}),
\end{equation}

where $\psi^{\ast}$ denotes the conjugate of the Morlet mother wavelet, and $\psi(t) = \pi^{-1/4} [e^{i \omega_{0} t} - e^{-\frac{\omega_{0}^{2}}{2}}] e^{-\frac{t^{2}}{2}}$. The parameter $w_{0}$, representing the frequency of the wavelet, is set equal to $w_{0}=6$. The transformation from the dilation scale, $\ell$, to the physical spacecraft frequency, $f_{sc}$, is given by:

\begin{equation}
f_{sc} = \frac{w_{0}}{2 \pi \ell \Delta t },
\end{equation}

where, $\Delta t$ represents the time interval between successive measurements. The power spectral density of the i-th component as a function of spacecraft frequency $f_{sc}$ and the local, scale-dependent field/flow angle $\theta_{BV}$ can be estimated as:

\begin{equation}
F_{ii}(f_{sc}, ~ \theta_{BV}) = \frac{2\delta t}{N} \sum_{n=0}^{N-1}|\omega_{i} (\ell, t_{n}, \theta_{BV})|^{2},
\end{equation}

Here, N is the number of samples within the range $\theta_{j-1} \leq \theta_{BV} \leq \theta_{j}$, $\theta_{j} = 5^{\circ} \cdot j$, j=0, 1, ..., 18. At time $t_n$ and wavelet scale $\ell$, we estimate the angle $\theta_{BV}$ using the scale-dependent local mean magnetic field $\boldsymbol{B}_{\ell}$ and velocity field $\boldsymbol{V}_{\ell}$, where $\boldsymbol{V}$ represents the solar wind velocity in the spacecraft frame \citep{Duan_2021, Cuesta_anisotropy}. To calculate the scale-dependent local mean of a field, $\boldsymbol{q}$, we use a Gaussian weighting scheme centered at $t_n$:

\begin{equation}
\boldsymbol{q}_{\ell}(t_{n}, \ell) = \sum_{m=0}^{N-1}\boldsymbol{q}_{m}~exp\left(-\frac{(t_{n}-t_{m})^{2}}{2\lambda^{2} \ell ^{2}}\right),
\end{equation}

where $\lambda$ is a dimensionless parameter that determines the scaling of the average. To ensure the robustness of our findings, we investigated two distinct values of $\lambda$, specifically $\lambda = 1$ and $\lambda = 3$. Remarkably, the results obtained for both cases were comparable, exhibiting differences in spectral exponents of only 0.01-0.02 \citep[see also][]{Gerick_2017}. The parameter $\theta_{BV}$ was determined  using two distinct methods: the non-scale dependent time-to-time velocity field value $\boldsymbol{V}(t)$ and the scale-dependent value, $\boldsymbol{V}_{\ell}$. Our results indicates that the outcomes obtained from both techniques are practically indistinguishable, which validates the minimal impact of interpolating $\boldsymbol{V}$ at the time points of $\boldsymbol{B}$ or only considering $\boldsymbol{V}(t)$ \citep[see also][]{Verdini_2018, WANG_2020}. Throughout the remainder of the study, we utilize $\boldsymbol{B}_{\ell}$ and $\boldsymbol{V}_{\ell}$ to estimate the $\theta_{BV}$ parameter considering the case where $\lambda = 3$. For intervals that are sampled at heliocentric distances greater than 0.5 AU and have a significant lack of plasma data, defined as having more than $10\%$ of solar wind velocity measurements missing, the angle $\theta_{BR}$ is used. This angle represents the angle between $\boldsymbol{B}_{\ell}$ and the scale-dependent radial component of the magnetic field, denoted as $B_{R\ell}$. To determine the reliability and consistency of using $\theta_{BR}$ instead of $\theta_{BV}$, both angles were evaluated for intervals with adequate plasma data. Our findings suggest that the anisotropic spectra remained almost unchanged for the majority of intervals, even when sampled as close as 0.3 au. The subsequent analysis examines the trace of the power spectral density, denoted as $F = \sum F_{ii}$. The range of $\theta_{BV}$ is restricted to be between $0^{\circ}$ and $90^{\circ}$ based on the symmetry of $\theta_{BV}$ around $90^{\circ}$ \citep{Chen-anisotropic_2011}.

To transform the PSD derived in the spacecraft-frame frequency $F(f_{sc}, ~ \theta_{BV})$ into a wavenumber spectrum expressed in physical units $E(\kappa^{\ast},~ \theta_{BV})$, we employ Taylor’s frozen-in hypothesis \citep{taylor_spectrum_1938}. This hypothesis assumes that the speeds of MHD wave modes, such as shear Alfv\'en modes propagating at $V_p=V_A\cos(<\boldsymbol k, \boldsymbol B>)$ observed in the solar wind plasma, are negligible compared to the bulk flow of the solar wind. This means that the Alfv\'en Mach Number $M_{A} = \frac{V_r}{V_A} \gg 1$. However, as the PSP spacecraft approaches the Sun, both the spacecraft velocity $\boldsymbol{V}_{sc}$ and the speeds of MHD wave modes start to become comparable to $\boldsymbol{V}_{sw}$ \citep{2021A&A...650A..22P}. For this reason, a modified version of Taylor's hypothesis, that accounts for wave propagation and spacecraft velocity, is adopted for heliocentric distances below 0.3 au, i.e., $\kappa^{\ast} = \kappa \cdot d_{i} = 2 \pi f_{sc}/V_{tot} \cdot d_{i}$, $d_{i}$ represents the ion inertial length, and $V_{tot} =| \boldsymbol{V}{sw} + \boldsymbol{V}{a} - \boldsymbol{V}_{sc}|$ \citep{Klein_2015, 2022_Zank, Zhao_2022ApJ}. It is important to note that this method assumes the dominance of outwardly propagating waves, which is the case for the vast majority of the analyzed intervals closer to the Sun \citep{2022ApJ...940L..13A, sioulas_turb_22_no1}.

\section{\label{sec:Data} Data Selection and Processing}

\subsection{Data Selection}\label{subsec:Data_selection}

As a first step, observations of PSP between January 1, 2018, and October 1, 2022, were collected, encompassing the first thirteen perihelia $(E1-E13)$ of the PSP mission. Level 2 magnetic field data from the Flux Gate Magnetometer (FGM) \citep{bale_fields_2016}, as well as Level 3 plasma moment data from the Solar Probe Cup (SPC) for E1-E8, and the Solar Probe Analyzer (SPAN) from the Solar Wind Electron, Alpha and Proton (SWEAP) suite for E9-E13 \citep{kasper_solar_2016}, were analyzed. Data from the SCaM product \citep{https://doi.org/10.1029/2020JA027813} obtained during E1 have also been analyzed and will be presented as a high-quality case study. The plasma data consists of moments of the distribution function computed on board the spacecraft, including the proton velocity vector $\boldsymbol{V}_{p}$, number density $n_p$, and temperature $T_p$. When available, electron number density data derived from the quasi-thermal noise from the FIELDS instrument \citep{Moncuquet_2020} were preferred over SPAN or SPC data for estimating proton number density. In order to calculate the proton density from the electron density, charge neutrality must be considered, leading to a $\approx4\%$ abundance of alpha particles. Therefore, electron density from QTN was divided by 1.08.


\begin{figure*}[htb!]
\centering
\setlength\fboxsep{0pt}
\setlength\fboxrule{0.0pt}

\includegraphics[width=1\textwidth]{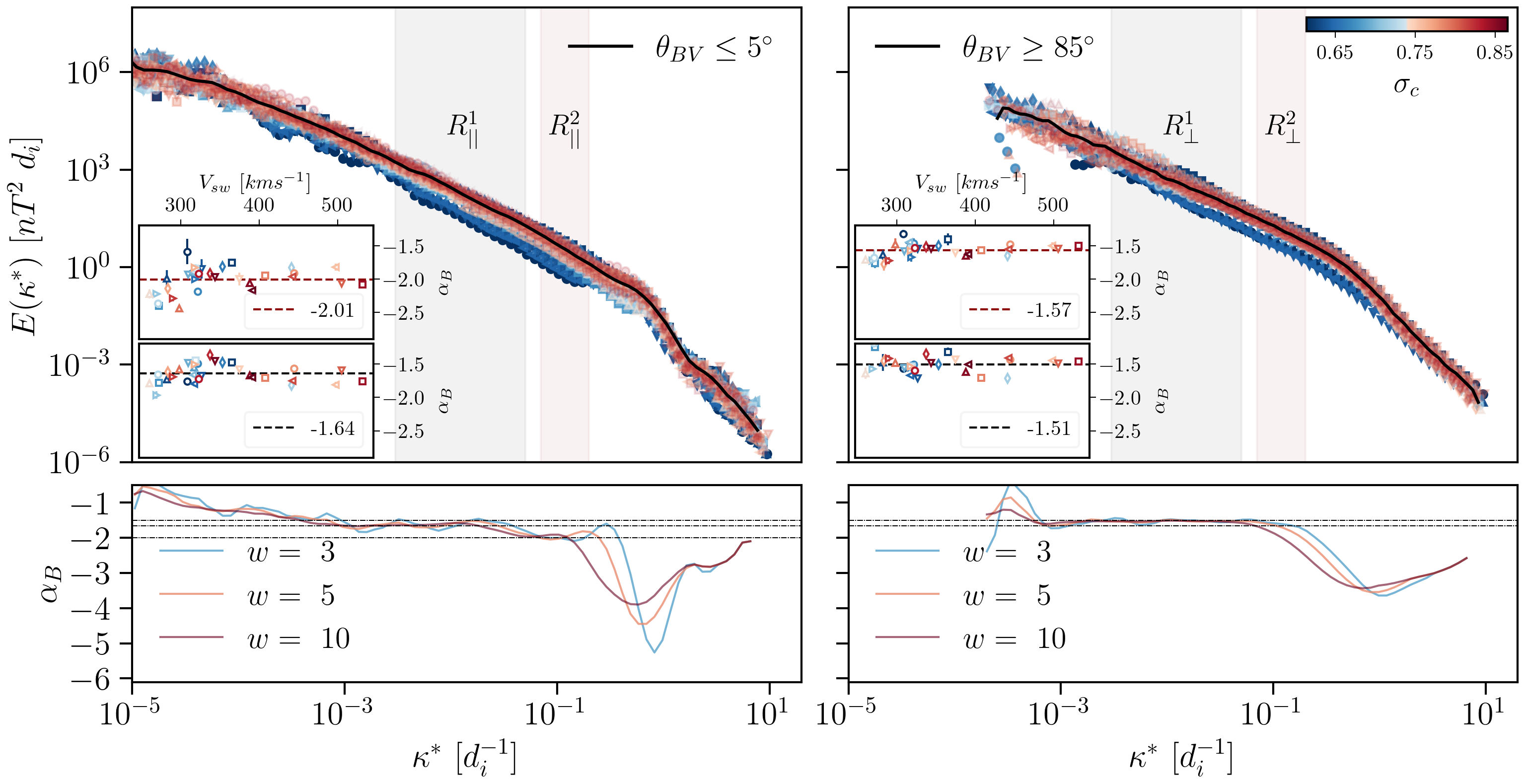}
\caption{Averaged magnetic field power-spectrum (black solid line) for fluctuations parallel  $\theta_{BV} \leq 5^{\circ}$ (left panel) and perpendicular $\theta_{BV} \geq 85^{\circ}$ (right panel) to the local magnetic field during the first perihelion of PSP (E1) estimated using SCaM data. The dependence of the spectra on normalized cross-helicity ($\sigma_c$) is also shown, with the color keyed to $\sigma_c$. The inset figures illustrate the spectral index $\alpha_{B}$, at two different ranges of scales (bottom) $3\times10^{-3} - 5 \times 10^{-2} d_i$ and (top) $8\times10^{-2} - 2 \times 10^{-1} d_i$ as a function of solar wind speed and $\sigma_c$. The dashed horizontal lines indicate the mean value of $\alpha_{B}$.
The second row illustrates the local $\alpha_{B}$, calculated over a sliding window of a factor of 3, 5, and 10 shown in cyan, orange, and red respectively. Horizontal  dotted lines have also been added marking values $-3/2$, $-5/3$ and $-2$}\label{fig:SCaM:E1_PSD}
\end{figure*}

\par
The second step involved obtaining magnetic field and particle measurements from the SO mission between June 1, 2018, to March 1, 2022. Magnetic field measurements from the Magnetometer (MAG) instrument \citep{horbury_solar_2020}. Note, that burst magnetic field data have been utilized when available. Particle moments measurements for our study are provided by the Proton and Alpha Particle Sensor (SWA-PAS) onboard the Solar Wind Analyser (SWA) suite of instruments \citep{owen_solar_2020}.

\subsection{Data processing}\label{subsec:Data_processing}

Quality flags for the magnetic field and particle time series have been taken into account, and time intervals missing $\geq 1\%$ and/or $\geq 10\%$ in the magnetic field and particle time series have been omitted from further analysis. Additionally, the mean value of the cadence between successive measurements $\delta \tau$ in the magnetic field time series has been estimated for each of the selected intervals, and time intervals that were found to have a mean cadence of $\delta \tau \geq 250$ ms were discarded. Due to poor data quality, all PSP intervals exceeding $R \simeq 0.5$ au have also been discarded.

Spurious spikes in the plasma time series were eliminated by replacing outliers exceeding three standard deviations within a moving average window covering 200 points with their median values \citep{davies_identification_1993}.

\section{Results}\label{sec:results}

\subsection{Case Study: SCaM Dataset, E1}\label{subsec:E1}

\begin{figure*}[htb!]
\centering
\setlength\fboxsep{0pt}
\setlength\fboxrule{0.0pt}

\includegraphics[width=1\textwidth]{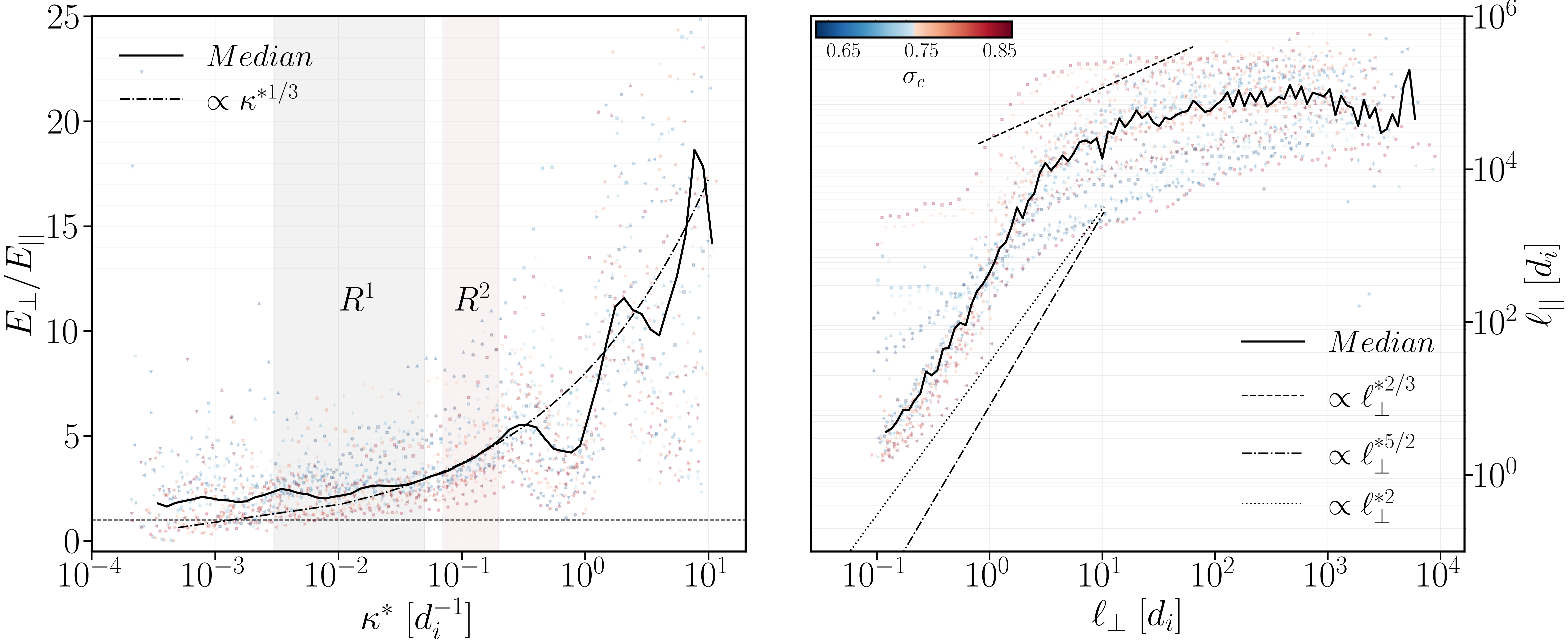}
\caption{(a) The anisotropy
of the fluctuations described by the ratio of the perpendicular ($E_{\perp}$) to
the parallel power  ($E_{||}$). A horizontal black solid line has been added to indicate $E_{\perp}/E_{||} =1$.  (b) The relation between $\ell_{||}$ and $\ell_{\perp}$, interpreted as wavevector anisotropy. In both cases,  intervals have also been binned based on their $\sigma_c$ value. The median curve for each bin is shown with the color of each curve keyed to  $\sigma_c$ }\label{fig:SCaM:E1_wavevector_power_anis}
\end{figure*}

The high-resolution data from the first perihelion of the PSP (R $\approx 0.17$ au) from November 1 to November 11, 2018, was analyzed. A total of 33 intervals each with a duration of 12 hours were obtained and the power spectral density was estimated, with subsequent intervals overlapping by $50\%$. The analysis covered 18 bins of the $\theta_{BV}$ angle, but in the following, we will only focus on those bins closer to the parallel and perpendicular directions, with $\theta_{BV} \leq 5^{\circ}$ and $\theta_{BV} \geq 85^{\circ}$, respectively. It is worth noting that the second half of E1 displayed significantly different characteristics compared to the first half, with the solar wind exhibiting higher speeds and a greater number of magnetic switchbacks \citep{2019Natur.576..237B}. It is well established that power spectra exhibit different characteristics when different solar wind speeds are considered due to the different types of fluctuations they transport \citep{Borovsky_2012}. Specifically, the fast solar wind is highly Alfv\'enic and characterized by large-amplitude, incompressible fluctuations, while the slow wind is generally populated by smaller amplitude, less Alfv\'enic, compressive fluctuations, that include convected coherent structures \citep{bruno_radial_2003,Matteini_2014, shi_alfvenic_2021, Sioulas_2022_intermittency, 2020ApJ...898..113Z}. Consequently, the spectral variation due to the differing plasma parameters of the selected streams was investigated. More specifically, we considered the dependence of the PSD on the solar wind speed, $V_{sw}$, the normalized cross helicity $\sigma_{c}$:

\begin{equation}
\sigma_{c} = \frac{E_{o} - E_{i}}{E_{o} + E_{i}},
\end{equation}

a measure of the relative amplitudes of inwardly and outwardly propagating Alfv\'en waves, and the normalized residual energy $\sigma_r$:

\begin{equation}
\sigma_{r} = \frac{E_{V} - E_{b}}{E_{V} + E_{b}},
\end{equation}

indicating the balance between kinetic and magnetic energy, where, $E_{\phi} = \frac{1}{2}\langle \delta \boldsymbol{\phi}^{2} \rangle$ denotes the energy associated with the fluctuations of the field $\boldsymbol{\phi}$. 
In particular, $E_{o,~ i}$ can be estimated using Elsasser variables, defining outward and inward propagating Alfv\'enic fluctuations \citep{Velli_91_waves, Velli_93} 
\begin{equation}
    \delta \boldsymbol{Z}_{o,i} = \delta\boldsymbol{V} \mp sign(B^{R}_{0})\delta \boldsymbol{b}, 
\end{equation}

$\delta \boldsymbol{B}  = \boldsymbol{B} - \boldsymbol{B_{0}}$, $\boldsymbol{B_{0}}$  the background magnetic field, $\delta \boldsymbol{b} = {\delta \boldsymbol{B}}/{\sqrt{\mu_0 m_p n_p}}$ the magnetic fluctuations in Alfv\'en units and $B^{r}_{0}$ the ensemble average of $B_{R}$, utilized to determine the polarity of the radial magnetic field  \citep{shi_alfvenic_2021}. The magnetic field power-spectrum for fluctuations parallel $\theta_{BV} \leq 5^{\circ}$ and perpendicular $\theta_{BV} \geq 85^{\circ}$ to the local magnetic field, resulting from averaging all the respective spectra are presented in Figure \ref{fig:SCaM:E1_PSD}.  Individual spectra are also shown with the color of the curve keyed to $\sigma_c$. The fluctuation power in the MHD range shows a positive correlation with $\sigma_c$, but this dependence vanishes in the transition region and kinetic scales. A similar trend was observed with $\sigma_r$ and $V_{sw}$, not shown here. The results are consistent with \citep{Vasquez_2007}, who found higher, MHD range, turbulence amplitudes associated with faster streams, as well as, \citep{Pi_2020} who showed that such dependence vanishes in the kinetic scales. The trend also vanishes at the large, energy injection scales,  $\kappa d_{i} \leq 10^{-3}$, where the power spectrum is clearly dominated by parallel fluctuations. Focusing our attention on MHD scales, we can observe two distinct ranges, roughly $3\times10^{-3} - 5 \times 10^{-2} \kappa d_i$ and $8\times10^{-2} - 2 \times 10^{-1} \kappa d_{i}$, over which the PSD displays a clear power-law scaling.
A light-black and red shade are used to indicate these regions in the figure, and we will thereby refer to them as $R_{|| (\perp)}^{1}$, and $R_{|| (\perp)}^{2}$. The power-law fitting has been applied to the PSD for the two ranges, and the bottom and top inset figures illustrate $\alpha_{B}$ as a function of $V_{sw}$. Note, that the color of the scatter plot is keyed to $\sigma_c$. Furthermore, horizontal lines have been added to indicate the average value of $\alpha_{B}$. In the direction parallel to the mean field the PSD scales roughly like -5/3 and -2 in $R_{||}^{1}$ and $R_{||}^{2}$, respectively. For perpendicular fluctuations, only a minor difference may be observed between $R_{\perp}^{1}$ and $R_{\perp}^{2}$ which are characterized by a power-law scaling with index -3/2 and -1.57, respectively. The absence of a definitive correlation between $\alpha_{B}$, $V_{sw}$, and $\sigma_{c}$, as reported in the study by \citep{sioulas_turb_22_no1}, could be ascribed to the relatively extended time intervals that were examined or the limited size of the sample, which, in this case, encompassed only 33 intervals.


\begin{figure*}[htb!]
\centering
\setlength\fboxsep{0pt}
\setlength\fboxrule{0.0pt}

\includegraphics[width=1\textwidth]{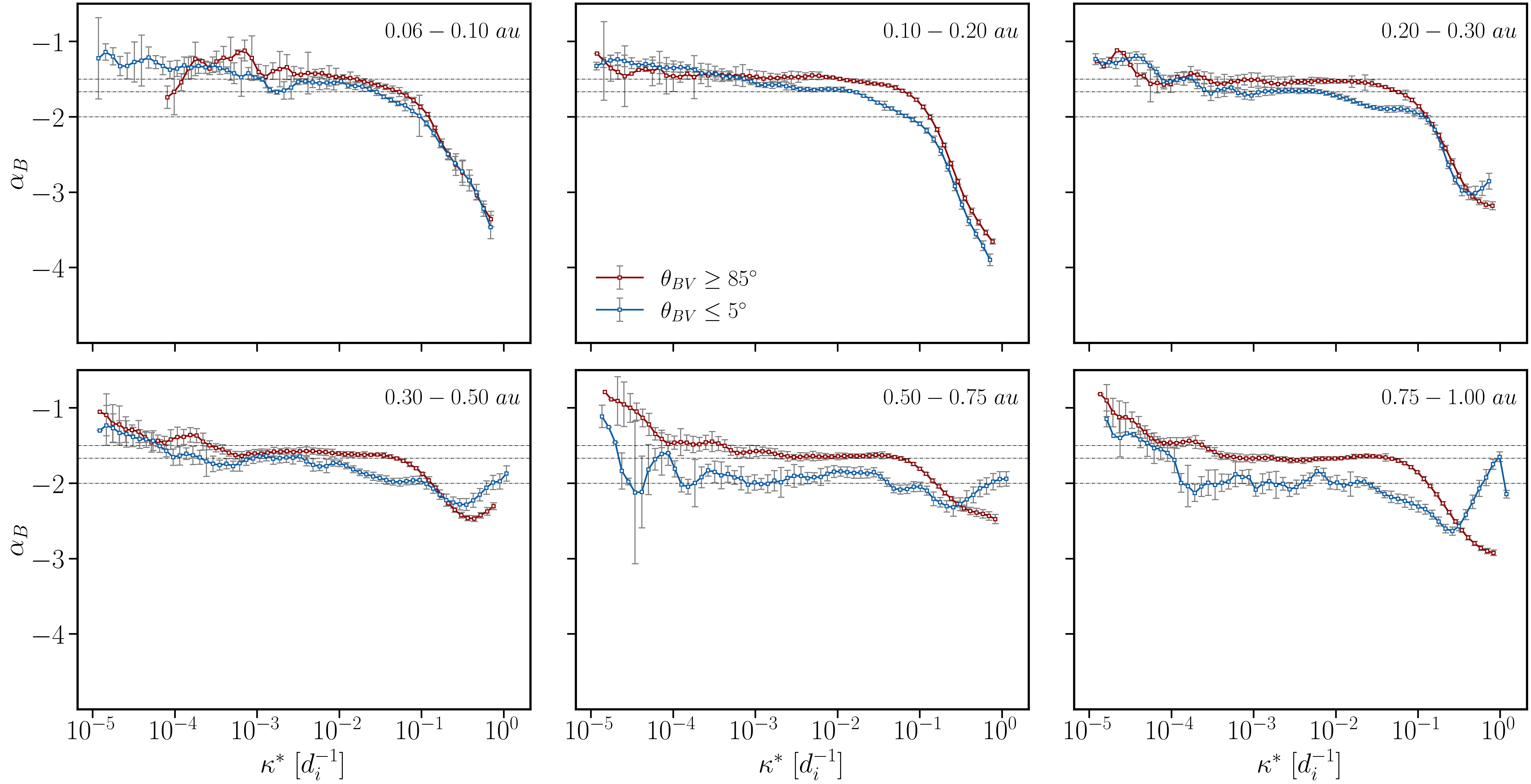}
\caption{ The local spectral index ($\alpha_{B}(\kappa^{\ast})$)
 for fluctuations with $\theta_{BV} \leq 5^{\circ}$ (blue),  and $\theta_{BV} \geq 85^{\circ}$ (red) at different heliocentric distances for slow streams, $V_{sw} \leq 400 km ~s^{-1}$. The local spectral index was calculated for all selected intervals, and each curve corresponds to the mean of all intervals that fall inside the bins indicated in the legend. We focus on MHD scales, $\kappa d_{i} \leq 3 \times 10^{-1}$, because instrumental noise flattens the PSD with increasing distance, as discussed in Section \ref{subsec:E1}. 
}\label{fig:alpha_evol_slow}
\end{figure*}
When examining the local spectral index, $\alpha_{B}(\kappa^{\ast})$, a similar pattern emerges. This is achieved by applying a sliding window of size $w = 3, ~ 5, ~10$ in $\kappa^{\ast}$ over the spectra and calculating the best-fit linear gradient in log-log space over this window, shown in cyan, orange, and red, respectively, in the bottom panel of Figure \ref{fig:SCaM:E1_PSD}. At smaller scales where $\kappa^{\ast}\geq 0.1d_{i}^{-1}$, both parallel and perpendicular fluctuations display a steeper spectrum between the inertial and kinetic ranges. The scaling behavior observed in the transition and kinetic ranges is consistent with the findings reported by \citep{Duan_2021}. Additionally, \citep{Duan_2021} report a scaling exponent of $-2$ for the parallel spectrum in the inertial range spanning $4 \times 10^{-1} - 2$ Hz, which corresponds to region $R_{||}^{2}$ in our analysis. It is worth noting, however, that $R_{||}^{2}$ does not encompass the entire inertial range. Specifically, $R_{||}^{1}$ covers most of the MHD range and is characterized by a shallower scaling exponent, $\alpha_{B} \approx -5/3$. The two different MHD scalings persist in most of the intervals studied, suggesting that this may be a consistent feature of the solar wind power spectrum in the vicinity of the Sun.

\begin{figure*}[htb!]
\centering
\setlength\fboxsep{0pt}
\setlength\fboxrule{0.0pt}

\includegraphics[width=1\textwidth]{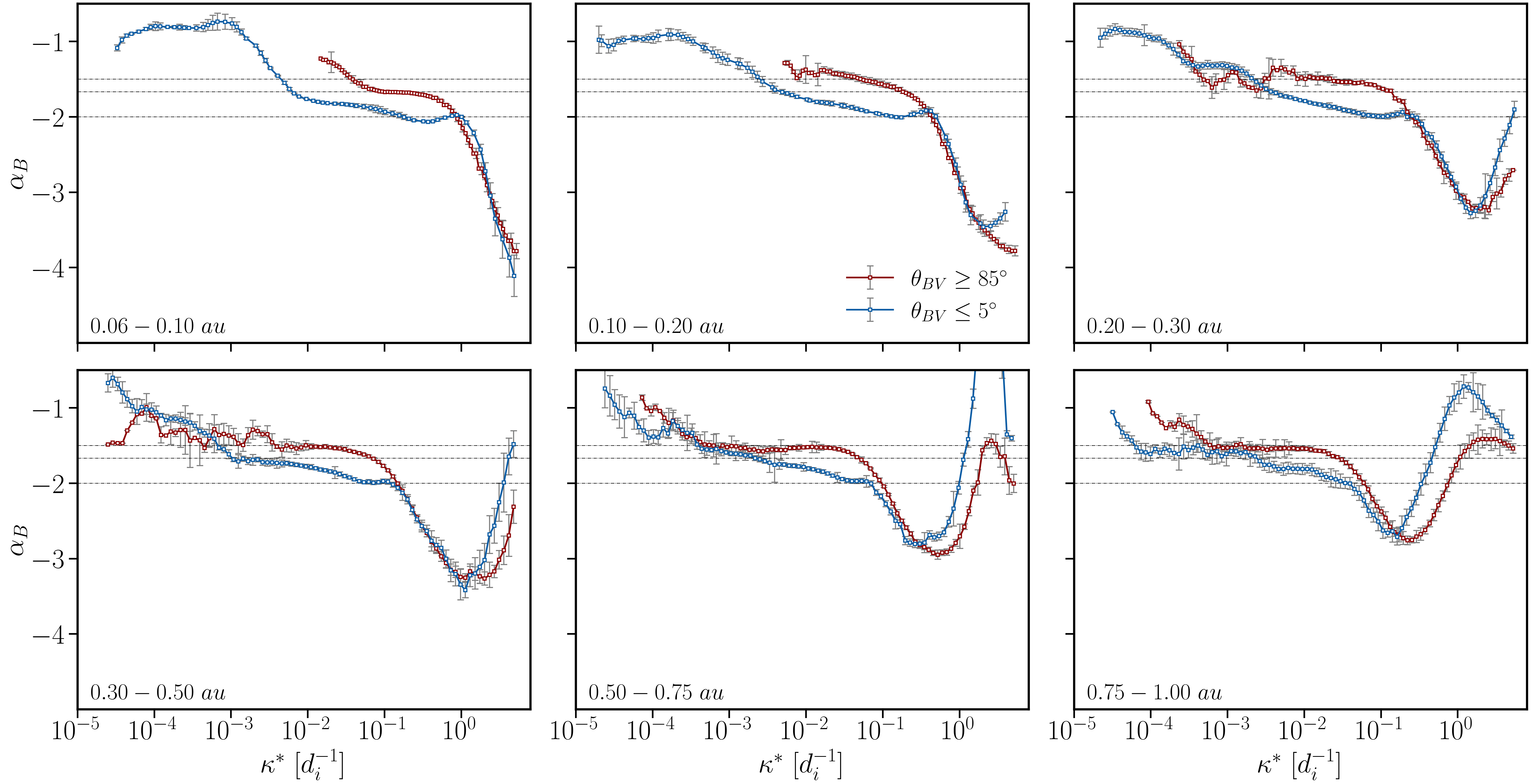}
\caption{ The local spectral index ($\alpha_{B}(\kappa^{\ast})$)
 for fluctuations with $\theta_{BV} \leq 5^{\circ}$ (blue),  and $\theta_{BV} \geq 85^{\circ}$ (red) at different heliocentric distances for fast streams, $V_{sw} \geq 400 km ~s^{-1}$. The local spectral index was obtained for all selected intervals and each curve corresponds to the mean of all the intervals that fall inside the bins indicated in the legend.
}\label{fig:alpha_evol_fast}
\end{figure*}

We then analyzed the power anisotropy, defined as $E_{\perp}/E_{||}$  \citep{Podesta_2009}, as a function of $\kappa^{\ast}$. The results of this analysis are displayed in Figure \ref{fig:SCaM:E1_wavevector_power_anis}, where individual intervals are plotted as scatter points and binned based on their $\sigma_c$ value. The median curve for each bin is shown and the color of each curve is keyed to $\sigma_c$. The median curve (black solid line) in Figure \ref{fig:SCaM:E1_wavevector_power_anis} is consistent with previous findings at larger heliocentric distances. Specifically, the curve exhibits a region of near isotropy for $kd_{i} \leq 10^{-3}$, which roughly corresponds to the roll-over to the $f^{-1}$ range of the magnetic spectrum (see Figure \ref{fig:SCaM:E1_PSD}). At smaller scales, the anisotropy becomes more noticeable and shows a power-law scaling that closely resembles the 1/3 value suggested by the CB conjecture. Therefore, a line with a scaling exponent of 1/3 was included in the figure as a point of reference. This $\kappa^{\ast 1/3}$ scaling is observed within the range of $4\times10^{-2} - ~3\times10^{-1} ~[d_{i}^{-1}]$, which corresponds to region $R^{2}$ in Figure \ref{fig:SCaM:E1_PSD}. Additionally, while the anisotropy increases at smaller scales until $\kappa d_{i} ~\approx 4 \times 10^{-1} $, there is a sudden but noticeable local minimum at around $\kappa d_{i} ~\approx 0.7$ followed by a local maximum at $\kappa d_{i} ~\approx 1.7$. Both the trough and peak are consistently observed across all intervals considered in this study. The local minimum may be caused by the bump observed in $E_{||}$ at $\kappa d_{i} \approx 0.06$, which coincides with the beginning of the transition region in $E_{\perp}$ (see Figure \ref{fig:SCaM:E1_PSD}). This bump may suggest a local enhancement of energy that could be due to ion kinetic instabilities \citep{wicks_anisotropy1}. For a more comprehensive discussion of the double-peak structure in Figure \ref{fig:SCaM:E1_wavevector_power_anis}a, see \citep{Podesta_2009}.The results of this study differ from those of \citep{Podesta_2009} in that we observe an increase in anisotropy at smaller scales $\kappa d_{i} > 2$. As shown in Figure 7 of \citep{Podesta_2009}, a rapid decrease in the power ratio is observed beyond the local kinetic scale maximum of approximately 1 Hz, which is attributed to the dissipation of kinetic Alfven waves (KAWs). However, as the spacecraft moves farther away from the sun, the amplitude of the fluctuations at kinetic scales is close to the noise floor of the magnetometer. This can lead to an artificial steepening of the power spectral density (PSD) caused by instrumental noise \citep{phdthesis}. The effect is particularly significant for $\alpha_{B}(\kappa^{\ast})$ parallel, as most of the power in the solar wind is associated with perpendicular fluctuations. As a result, the parallel PSD systematically obtains lower values at MHD and kinetic scales and is therefore more likely to be affected by instrumental noise. On the other hand, the perpendicular PSD can remain intact. This can cause the parallel PSD to flatten out and the power ratio to decrease with decreasing scale. Considering that (1) the aforementioned paper uses magnetic field data from the STEREO mission \citep{Stereo_magnetometer} at Earth-orbit, where the turbulence amplitude is lower compared to that observed by PSP's E1, and (2) the SCaM data product merges fluxgate and search-coil magnetometer measurements, allowing for magnetic field observations up to 1 MHz with an optimal signal-to-noise ratio, we attribute the discrepancy to instrumental noise that may have affected the parallel PSD in Figure 7 of \citep{Podesta_2009}.

To estimate the anisotropy relation between $\ell_{||}$ and $\ell_{\perp}$, we equate the second-order structure functions for parallel and perpendicular fluctuations, $SF^{2}_{ ||}$ and $SF^{2}_{\perp}$, respectively, estimated as

\begin{equation}
SF^{2} (\ell^{\ast}, \theta_{BV}) = \frac{(d_{i}/V_{SW})}{N}\sum_{n=1}^{N}|\frac{\boldsymbol{\omega}(t_{n}, \ell, \theta_{BV})}{\sqrt{\tau}}|^{2}.
\end{equation} \label{eqn:Sfq Definition}

The resulting anisotropy is shown in Figure \ref{fig:SCaM:E1_wavevector_power_anis}b. In the region of $R^{2}$, the anisotropy follows a power-law scaling that is close to 2/3, meaning that the scaling is in rough accordance with CB, indicating that magnetic fluctuations are elongated along the magnetic field. The anisotropy becomes even stronger at smaller scales, where scalings of approximately 5/2 and 2 are observed in the transition and kinetic range, respectively. 

\par

\begin{figure*}[htb!]
\centering
\setlength\fboxsep{0pt}
\setlength\fboxrule{0.0pt}

\includegraphics[width=1\textwidth]{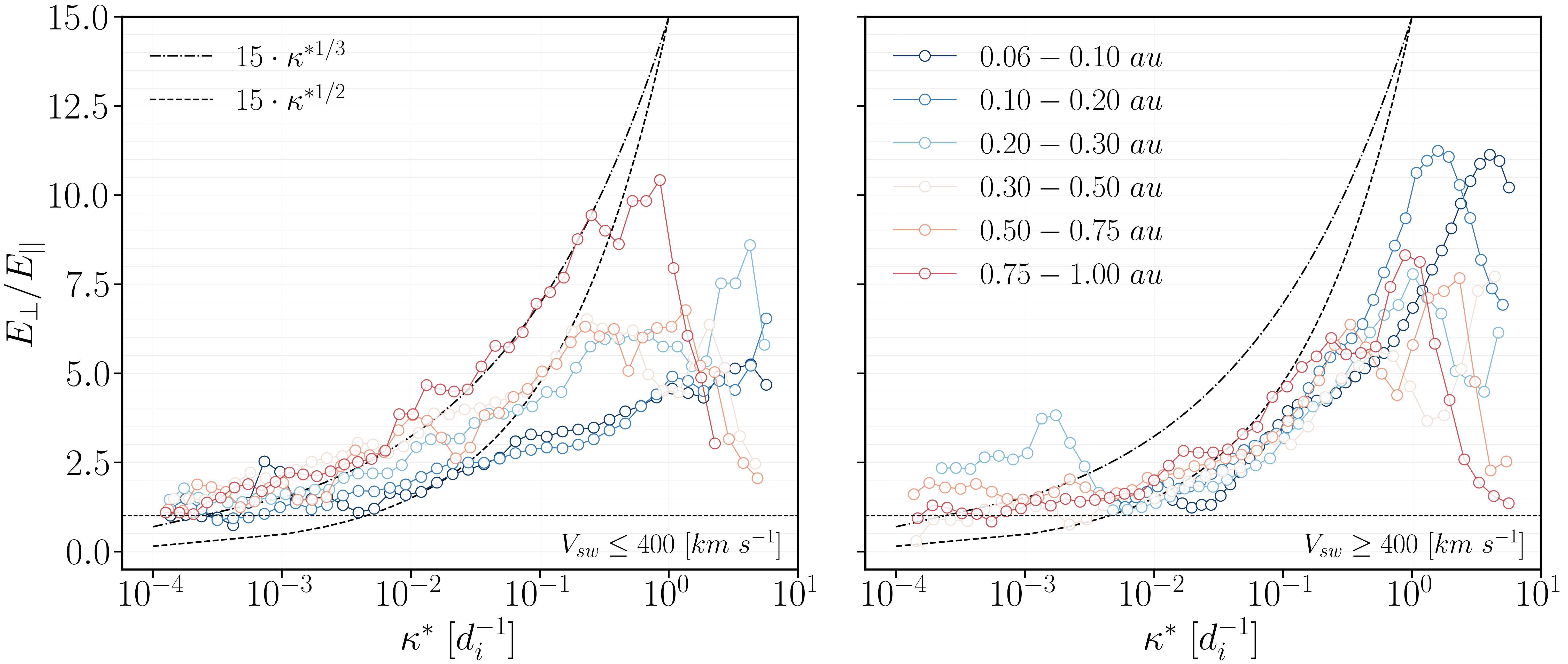}
\caption{The radial evolution of the power anisotropy, represented by the ratio $E_{\perp}/E_{||}$, is depicted as a function of heliocentric distance for slow ($V_{sw} \leq 400 ~km ~s^{-1}$) and fast ($V_{sw} \geq 400 ~km ~s^{-1}$) streams. Six heliocentric-radius bins were utilized, and each curve represents the median of $E_{\perp}/E_{||}$ from all intervals that fall within each bin. The dashed ($y= 15 \cdot \kappa^{\ast 1/2}$) and dashed-dotted ($y= 15 \cdot \kappa^{\ast 1/3}$) lines were added as reference points, indicating consistency with ``dynamically aligned" and ``critically balanced" cascade, respectively. }\label{fig:power_anis_evolution}
\end{figure*}

\subsection{Radial Evolution of  anisotropic turbulence}\label{subsec:radial_evolution_of_turbulence}

\subsubsection{Spectral-index anisotropy}\label{subsubsec:pectral-index anisotropy}

In the following, we investigate the evolution of spectral index anisotropy with heliocentric distance. For this analysis we consider intervals sampled by the PSP and SO at distances between 0.06 - 1 au (see Section \ref{subsec:Data_processing}). Previous research has shown that the dominant orientation of fluctuation wavevectors in fast solar wind streams tends to be quasi-parallel to the local magnetic field, while in slow solar wind streams the dominant orientation is quasi-perpendicular \citep{Dasso_2005}. In order to examine the distinct features of each type of stream and their potential impact on the development of anisotropy in the solar wind, a comprehensive visual analysis was undertaken to categorize the streams into two distinct groups: slow streams characterized by $V_{sw} \leq 400 ~km/s$ and fast streams with $V_{sw} \geq 400 ~km/s$ . A comprehensive record of the chosen intervals can be accessed in  \href{https://github.com/nsioulas/MHDTurbPy}{\textcolor{blue}{MHDTurbPy}}

We shall begin by examining the evolution of slow streams, which comprise the majority of the samples collected from PSP and SO. To determine the local spectral index for each interval, we perform calculations in the direction parallel ($\theta_{BV} \leq 5^{\circ}$) and perpendicular ($\theta_{BV} \geq 85^{\circ}$) to the locally dominant magnetic field, utilizing a sliding window of size $w = 10$, following the methodology outlined in Section \ref{subsec:E1}. 

We then partitioned our intervals into six heliocentric bins and calculated the mean local spectral index for those intervals that fell within each bin. It should be noted that despite the spectra and local spectral indices being calculated at identical frequencies based on the interval duration and sampling frequency, the normalization process results in an irregular shift along the vertical axis. Consequently, we divided the complete range of $\kappa^{\ast}$ into 100 bins, and computed the mean for all $\alpha_{B}(\kappa^{\ast})$ values that fell within each bin, as described in \citep{Nemecek_2021}. It is worth noting that the size of the interval under consideration does not have a significant impact on the outcomes. This is true as long as a sizable statistical sample of fluctuations is taken into account at a given scale, in order to ensure the validity of the statistical analysis and produce accurate spectra \citep{Dudok_de_wit_Samples_Rule}. Any intervals that exhibited noisy or otherwise unreliable spectra were excluded from subsequent analyses. However, it is important to note that such intervals made up only an inconsequential proportion of the overall dataset.

The radial size of each bin is shown in the legend of Figure \ref{fig:alpha_evol_slow}. The slow wind local spectral indices, as a function of heliocentric distance, are shown in blue for perpendicular fluctuations, and in red for parallel fluctuations along with error bars indicating the standard error of the mean. The standard error is given by $\sigma_{i}/ \sqrt{N}$, where $\sigma_{i}$ is the standard deviation and $N$ is the number of samples inside the bin, as described in \citep{10.2307/2682923}. We focus our attention on MHD scales, $\kappa d_{i} \leq 3 \times 10^{-1}$, here simply because instrumental noise artificially steepens the PSD with increasing distance, as discussed in Section \ref{subsec:E1}.

It is evident from Figure \ref{fig:alpha_evol_slow} that the spectral-index anisotropy of slow wind turbulence diminishes closer to the Sun. Within 0.1 au, both parallel and perpendicular spectra are characterized by a poorly developed inertial range, viz. the range of scales over which the spectral index is constant is limited to $3 \times 10^{-3} \lesssim \kappa d_{i} \lesssim 2 \times 10^{-2}$ with a scaling exponent  $-1.47 \pm 0.04$ and  $-1.55 \pm 0.05 $ for perpendicular and parallel fluctuations. At distances 0.1-0.2 au, two subranges ($R^{1}$ and $R^{2}$) emerge within the inertial range. The transition occurs at $\kappa d_{i} \approx 6 \times 10^{-2}$, and the scaling exponents in these ranges are similar to those shown in Figure \ref{fig:SCaM:E1_PSD}. However, the steepened region $R^{2}$ is not as well defined in this case. Considering that the PSP was at a distance of 0.17 au during E1, we attribute this discrepancy to the fact that $R^{2}$ actually appears closer to 0.2 au. By shifting the left boundary of the bin towards 0.2 au, we confirmed this expectation as the steepened region displayed a parallel power-law scaling of -2 when the left boundary was shifted to approximately 0.15 au.

In $R^{1}$ both parallel and perpendicular spectra dynamically evolve with increasing distance and steepen towards -5/3, which in the case of the parallel spectrum occurs within 0.1 au. The steepening occurs in a scale-dependent fashion which results in $R^{2}$ extending to larger scales with distance. As a result, for distances exceeding 0.5 au, $R^{1}$ practically vanishes, and the power spectra are characterized by a power-law exponent that changes from $-5/3$ in the direction perpendicular to $-2$ in the direction parallel to the locally dominant mean field in good agreement with the predictions of ``critical balance'' theory. It should be noted that the analysis was iterated over bins of width $10^{\circ}$, yielding consistent outcomes. Notably, the obtained Power Spectral Densities (PSDs) for $\theta_{BV} \geq 80^{\circ}$ or $\theta_{BV} \geq 85^{\circ}$ exhibited indistinguishable scaling behavior across all distances. Conversely, a comparison of the PSDs obtained for $\theta_{BV} \leq 10^{\circ}$ and $\theta_{BV} \geq 5^{\circ}$ revealed marginally steeper scaling behavior in the latter case for the inertial range. Specifically, in the instance of slow solar wind intervals, when distances exceeded 0.5au, a consistent -2 scaling was observed for $\theta_{BV} \geq 5^{\circ}$, whereas for $\theta_{BV} \geq 10^{\circ}$, the scaling behavior obtained was closer to -1.89.

We next examined the evolution of fast streams ($V_{sw} \geq 400 ~ km~ s^{-1}$). It is important to consider that as the solar wind expands in the heliosphere, the local mean magnetic field vectors become increasingly oriented at larger angles relative to the radial direction. This radial trend causes sampling at 0.06 AU to be more quasi-longitudinal, and sampling at 1.0 AU to be more quasi-perpendicular. As parallel fluctuations decrease with increasing distance, our ability to accurately estimate the low-frequency part of the parallel power spectrum is reduced. This effect makes the determination of the anisotropic scaling laws for high-speed streams in the ecliptic plane challenging, as there is insufficient data to make accurate measurements at low frequencies. While using longer records could resolve this issue, the limited lifetime of the streams restricts the length of the record. In an effort to address this issue, we imposed a minimum interval length that would allow for a large enough interval size but still enable us to gather a sufficient number of intervals for our statistical study. Specifically, for heliographic distances exceeding 0.3, and 0.5 au, we set the minimum interval size to 12 and 20 hours respectively. This resulted in a total of 274 intervals sampled across the inner heliosphere. The results of this analysis are presented in Figure \ref{fig:alpha_evol_fast}. It is readily seen that the differences between fast and slow intervals are significant. When examining the lower frequencies, we observe that within 0.2 au, the energy injection range of the PSD is dominated by parallel fluctuations. In particular, a remarkably extended and relatively shallow range with $\alpha_{B} \approx -0.8$ is observed within 0.1 au, which steepens towards -1 with distance. This is particularly noteworthy as previous research has shown that Alfvén waves (AWs) can parametrically decay into slow magnetosonic waves and counter-propagating AWs  \citep{Galeev_1963_PDI, Tenerani_2017ApJ_PDI, Malara_22_PDI}. This process may lead to the development of a $k_{||}^{-1}$ spectrum for outward-propagating AWs by the time they reach a heliocentric distance of 0.3 au in the fast solar wind \citep{chandran_2018}. For a more comprehensive investigation of the radial evolution of the lower-frequency part of the spectrum, see \citep[][, submitted to APJL]{davis2023, huang2023}. Due to the issues with interval size that were discussed earlier, we do not attempt to interpret the evolution of the lower-frequency part of the spectrum beyond 0.3 au.\par

Focusing on MHD scales, we notice that the perpendicular PSD only extends up to $\kappa d_i \approx 10^{-2}$. This implies that fast streams in proximity to the Sun exhibit a nearly radial magnetic field at low frequencies. Interestingly, within 0.1 au, the scaling of the perpendicular spectrum is consistent with $-5/3$, but at larger distances, a scaling that is roughly consistent with $-3/2$, fluctuating between -1.49 to -1.55, is observed. This suggests that the MHD range spectral index of the perpendicular spectrum for fast streams may not evolve in a consistent manner with increasing distance in the inner heliosphere. It is worth noting, however, that within 0.1 au, only four intervals with $V_{sw}\geq 400$ km/s were sampled by PSP. More data from fast streams near the Sun is needed to statistically confirm these findings. For parallel fluctuations, the inertial range scaling remains remarkably similar across all heliographic bins with the spectral index progressively steepening towards smaller scales from -5/3 towards -2, where a narrow range of scales over which the local spectral index obtains a constant value appears. In contrast to slow wind streams, the high-frequency point in fast wind streams does not remain anchored in a normalized wavenumber but gradually drifts towards larger scales with distance. This is an interesting finding that suggests the evolution of the high-frequency point is different between fast and slow wind streams and is discussed further in Section \ref{sec:discussion}.

\subsubsection{Power anisotropy}\label{subsubsec:power anisotropy}

In this section we examine the radial evolution of the power anisotropy, represented by the ratio $E_{\perp}/E_{||}$, where $E_{\perp}, ~ E_{||}$ the PSD for $\theta_{BV} \geq 85^{\circ}$ and $\theta_{BV} \leq 5^{\circ}$, respectively. To do this, we utilized the method described in Section \ref{subsubsec:pectral-index anisotropy} and calculated the mean of $E_{\perp}/E_{||}$ in six heliocentric bins. The results of this analysis are presented in Figure \ref{fig:power_anis_evolution}a for slow streams and Figure \ref{fig:power_anis_evolution}b for fast streams. According to theories based on ``dynamical alignment', the inertial range scaling index should be 1/2 when considering $E_{\perp}/E_{||}$, while a slope of 1/3 is predicted by theories of ``critical balance''.

For slow wind streams, the power anisotropy becomes more significant with increasing distance, particularly at smaller scales (see Figure \ref{fig:power_anis_evolution}a). This suggests that the turbulence undergoes an anisotropic cascade, transporting the majority of its magnetic energy towards larger perpendicular wavenumbers. In contrast, fast streams show practically no significant radial trend, especially when taking into account the error bars (not shown here). As a result, even though the power anisotropy is more pronounced for fast winds closer to the Sun, at distances of around 1 au, the situation is reversed, and slow wind exhibits higher values of $E_{\perp}/E_{||}$. In terms of anisotropic scaling, we observe that $E_{\perp}/E_{||}$ evolves in a manner similar to what was described in Subsection \ref{subsubsec:pectral-index anisotropy}. Specifically, the scaling of $E_{\perp}/E_{||}$ for slow wind streams does not fit the predictions of any of the existing anisotropic theories closer to the Sun, but with increasing distance, it evolves towards a scaling that is consistent with CB theories. The situation is more complex for fast streams. In particular, for the bin closest to the Sun, the scaling of $E_{\perp}/E_{||}$ is closer to that predicted by CB theories ($\kappa^{\ast 1/3}$), but for the rest of the bins, the scaling exponent fluctuates in the range between $1/2~-~1/3$. Additionally, the double peak structure discussed in Section \ref{subsec:E1} is also observed for most of the curves in this analysis, especially in fast wind intervals. In contrast to the data presented in Section \ref{subsec:E1}, at smaller scales, the utilization of fluxgate magnetometer data leads to a significant impact of instrumental noise on the resulting curves, ultimately causing a marked decrease in the power ratio.

\section{Discussion }\label{sec:discussion}

Wavelet analysis of solar wind data obtained at heliocentric distances greater than 0.3 au has shown strong agreement between the anisotropic characteristics of magnetic turbulence and the predictions of the ``critical balance'' conjecture \citep{horbury_anisotropic_2008, wicks_anisotropy1}. However, \citep{Podesta_2009} cautioned that it would be premature to draw conclusions about the agreement of the scaling in the fast solar wind with any particular theory due to the large uncertainties of the scaling at the largest scales. It is worth noting that these studies either focused on high-speed streams or prolonged periods of both high-speed and slow streams\citep{horbury_anisotropic_2008, wicks_anisotropy1, wicks_alignmene_2013, He_2013}. When extended intervals are considered, the PSD behavior will be practically determined by the fast sub-intervals since high-speed streams exhibit higher-amplitude magnetic fluctuations. Recent PSP measurements below 0.3 au have provided an unprecedented opportunity to study the nature of the solar wind in the vicinity of the solar wind sources. \citep{Bandyopadhyay_2021, Adhikari_2022} have recently shown that large-scale fluctuations in the near-Sun solar wind are dominated by wavevectors quasi-parallel to the local magnetic field.  \citep{Zhao_2022ApJ} also studied the radial dependence of this ratio by grouping the available datasets into two catalogs according to the radial distance and found that the ratio between parallel and perpendicular fluctuations observed by PSP is about $50\%:50\%$.


Inertial range spectral anisotropy has been investigated by \citep{Huang_2022, Wu2022OnTS}, who used slow solar wind data from E1 of PSP to show that the spectral indices are close to $-5/3$ and $-3/2$ in the parallel and perpendicular direction, respectively. \citep{Wu2022OnTS} furhter conducted a comparative analysis of the anisotropic spectral properties of the slow wind stream observed by PSP during E1 and a fast wind stream with a solar wind speed  $V_{sw} \approx 770 km/s$, which was sampled by Ulysses at 1.48 au. Their analysis led to the conclusion that the dynamical evolution of the inertial range scaling can be attributed to the existence of two sub-ranges in the inertial range.
Specifically, the sub-range closer to the kinetic scales, 30-300 $d_i$, exhibits a radial steepening, while the sub-range at larger scales remained unchanged.  As demonstrated by \citep{Wu2022OnTS}, the transition between the two ranges in question does not exhibit radial evolution, but rather remains constant in terms of $\kappa d_{i}$. Nevertheless, the reliability of this finding is uncertain given the contrasting radial evolutions of fast and slow streams based on turbulence signatures, as reported by \citep{shi_alfvenic_2021, Sioulas_2022_intermittency, Sioulas_2022_spectral_evolution}.

There are several significant questions that remain unanswered regarding the anisotropy of magnetic turbulence in the solar wind and its evolution as it propagates into the heliosphere. Firstly, it is unclear whether the anisotropy dynamically evolves with distance. Secondly, there is a need to investigate potential differences in spectral and power anisotropy between fast and slow streams, and if such differences exist, it is important to determine whether they evolve with distance. In the subsequent section, we endeavor to address these outstanding issues by comparing our findings with those of previous studies.

\section{Comparison with prior investigations. }\label{sec:Comparison}

\subsection{\cite{horbury_anisotropic_2008} $\&$ \cite{Podesta_2009} }\label{subsec:Comparison_Horbury_Podesta}

In our study, using data from PSP and SO, we estimated perpendicular inertial range spectral indices for the fast wind with values in the range of $ [-1.49, ~ -1.55]$. These values are slightly shallower than those reported in previous studies \cite{horbury_anisotropic_2008, Podesta_2009, wicks_anisotropy1} , which estimate values in the range of $ [-1.55, ~ -1.67]$. One possible reason for this discrepancy could be that the PSP and SO data were only collected in the ecliptic plane during the minimum and early rising phase of the Solar Cycle. It is known that solar wind conditions can vary significantly over the course of the Solar cycle, and it is possible that these variations could affect the observed scalings of the perpendicular spectra. In addition, due to the phase of the Solar Cycle, only a limited number of extended fast wind streams were collected. For example, PSP only sampled four intervals with $V_{sw}\geq 400$ km/s within 0.1 au. This limitation may affect the statistical significance of the results and make it difficult to accurately measure the anisotropic scaling laws for these streams at lower frequencies. As a result, it may be premature to draw firm conclusions about the agreement of the scaling in the fast solar wind sampled in the ecliptic plane by PSP and SO with any particular theory of anisotropic MHD turbulence.

\subsection{\cite{wicks_anisotropy1} }\label{subsec:Comparison_Wicks}

Our results indicate that, when analyzing slow wind streams, normalizing the PSD with $d_{i}$ allows us to fix the high-frequency break point, $f_{b}$, in normalized wavenumber space, as previously reported in \citep{sioulas_turb_22_no1}.  However, for fast solar wind streams, $f_{b}$ tends to shift towards larger $\kappa d_{i}$ as the distance increases. This phenomenon can be attributed to the fact that fast solar wind streams are characterized by higher proton temperatures ($T_{p}$) \citep{2020ApJS..246...62M, shi_alfvenic_2021,2023_Shi}, which lead to higher plasma pressure and, consequently, higher plasma $\beta$ values. The plasma $\beta$ is defined as the ratio of thermal to magnetic pressure, $\beta \equiv n_{p} K_{B} T_{p} / (B^{2}/2 \mu_{0})$, and it is comparatively higher in fast streams than in slow ones. It should be noted that the $V_{sw} - \beta$ correlation was verified, although it is not presented in this report. \citep{2014GeoRL..41.8081C, 2018_vech} have shown that the $f_{b}$ of the magnetic PSD between inertial and kinetic scales correlates better with $d_{i}$ when the intervals are characterized by $\beta <1$ values, while high $\beta$ intervals are characterized by a small scale break at the thermal ion gyroradius ($\rho_{i}$). In line with this, our analysis confirms the findings of \citep{wicks_anisotropy1}, who used five fast solar wind streams with $\beta >1$ between 1.5 - 2.8 au and found that the small scale end of the inertial range seems to naturally scale with the ion gyroradius when normalized with $\rho_{i}$. Given that $\rho_{i}$ grows radially as $\propto R^{1.48 \pm 0.02}$ \citep{sioulas_turb_22_no1}, we expect that $f_{b}$ will display a similar radial trend for fast solar wind streams.

\subsection{\cite{Wu2022OnTS} }\label{subsec:Comparison_Wu}

The use of high-resolution data from E1 of PSP allowed us to confirm the existence of two sub-ranges \citep{telloni_transition, Wu2022OnTS} within the inertial range. The transition occurs at $\kappa d_{i} \approx 6 \times 10^{-2}$ and signifies a shift from -5/3 to -2 scaling in the parallel spectra and from -3/2 to -1.57 scaling in the perpendicular spectra. The difference between the two ranges ($R^{1},~ R^{2} $) is most apparent in the parallel spectrum and could signify a transition from weak to strong turbulence\citep{1994_goldreich, Meyrand_transition, Zank_2020_transition}. It is important to note that the parallel spectral index we report here for $R_{||}^{2}$, $\alpha_{B} \approx -2$, is steeper than the one reported by \cite{Wu2022OnTS}. It is unlikely that the variations seen in the outcomes are due to the utilization of structure functions in the analysis carried out by \cite{Wu2022OnTS}. This is because we used second-order structure functions to confirm the anisotropic scaling. Nonetheless, it is feasible that the differences could be linked to the usage of better quality SCaM data in our research. In particular, Wu et al.'s parallel structure function in Figure 4b seems to become steeper at shorter timescales, but the limited cadence of around 1 Hz might prevent the clear detection of such scaling. \par

Moreover, it has been observed that there exist noteworthy differences between fast and slow streams in terms of their anisotropic properties and dynamic evolution. Besides the distinctions noted in the evolution of the inertial range scaling, the high-frequency breakpoint displays a more rapid shift towards lower frequencies in the analysis of fast wind streams. These findings are at odds with the assertions put forth by \cite{Wu2022OnTS} and emphasize the necessity of analyzing wind streams of comparable speeds for making meaningful comparisons.

\par

\section{CONCLUSIONS \&  SUMMARY
}\label{sec:Conclusions}

We used a merged PSP and SO dataset to study the dynamic evolution of turbulence anisotropy in the inner heliosphere, with a focus on understanding the differences in anisotropy observed between fast and slow wind streams

 \vspace{1em}
The main findings of our study can be summarized as follows:

    \vspace{1em}

    For slow wind streams , $V_{sw} \leq ~ 400 ~km ~s^{-1}$, we find:

    \vspace{0.5em}
    (a1) Within 0.1 au, the spectral index anisotropy of the inertial range vanishes, and the inertial range is confined to $3 \times 10^{-3} \lesssim \kappa d_{i} \lesssim 2 \times 10^{-2}$. The scaling exponents are $-1.47 \pm 0.04$ and $-1.55 \pm 0.05$ for perpendicular and parallel fluctuations, respectively. The power anisotropy ($E_{\perp}/ E_{||}$) is weaker compared to previous studies at 1 AU, and its inertial range scaling does not fit any predictions of anisotropic theories of turbulence.

    \vspace{0.3em}
    (a2) At $\approx 0.15$ au two inertial subranges ($R^{1}, ~ R^{2}$) emerge. The transition occurs at $\kappa d_{i} \approx 6 \times 10^{-2}$, and signifies a shift from -5/3 to -2 and -3/2 to $\approx-1.6$ scaling in parallel and perpendicular spectra, respectively. 

    \vspace{0.3em}
    (a3) Beyond this point, the power anisotropy monotonically strengthens with distance, indicating an anisotropic turbulent cascade that transports most of its magnetic energy towards larger perpendicular wavenumbers. Additionally, region $R^{2}$ extends towards smaller wavenumbers, gradually "consuming" region $R^{1}$. This process results in a scale-dependent steepening of the inertial range.

    \vspace{0.3em}
    (a4) At distances exceeding 0.5 au, region $R^{1}$ practically vanishes, and the power spectra are characterized by a power-law exponent that changes from $-5/3$ in the direction perpendicular to $-2$ in the direction parallel to the locally dominant mean field in good agreement with the predictions of ``critical balance''.

    \vspace{0.3em}

    (a5) The rate at which the high-frequency breakpoint $f_{b}$ of the magnetic power spectrum drifts to lower frequencies with distance scales naturally with the rate at which the ion inertial scale, $d_i$, grows with distance. In other words, the high-frequency point $f_{b}$ is observed to remain anchored in $\kappa d_{i}$.

    \vspace{1em}

   For fast streams, $V_{sw} \geq ~400 km s^{-1}$, we find:

   \vspace{0.5em}
   (b1)  Closer to the Sun, the energy injection range, $\kappa d_{i} \leq 10^{-3}$, of the spectrum is dominated by parallel fluctuations. Within 0.1 AU, this range exhibits a quite extended shallow region with a scaling index of $\approx -0.8$. This region appears to steepen towards -1 with increasing distance, providing evidence for the Parametric decay instability (PDI) as as generating mechanism for the $k_{||}^{-1}$ spectrum  in the fast solar wind \citep{chandran_2018}.
   
    \vspace{0.3em}
   (b2) In MHD scales, the scaling of both the parallel and perpendicular spectra does not exhibit a clear radial trend. 
   Within 0.1 au, the scaling of the perpendicular spectrum is consistent with $-5/3$. Beyond 0.1 AU, the perpendicular spectral index fluctuates between -1.49 and -1.55. For parallel fluctuations, the inertial range scaling remains remarkably similar across all heliographic bins. The spectral index progressively steepens towards smaller scales from -5/3 towards -2, where a narrow range of scales over which the local spectral index obtains a constant value is observed.

    \vspace{0.3em}
   (b3) Power-anisotropy for fast streams does not seem to display a clear trend with distance. In terms of inertial range scaling, we find that fast streams are more consistent with \citep{boldyrev_2006} model based on ``dynamical alignment'' than\citep{goldreich_magnetohydrodynamic_1997} model based on ``critical balance''  but the large uncertainties at lower frequencies make the statistical significance of this result questionable.
   
   \vspace{0.3em}
    (b4) In agreement with \citep{wicks_anisotropy1}, the high-frequency point, $f_{b}$ is observed to remain anchored in $\kappa \rho_{i}$.

\vspace{1em}

\par

A deeper understanding of anisotropy could be gained by considering the effect of intermittency on turbulence \citep{oboukhov_1962}, i.e., the concentration of fluctuation energy into smaller volumes of space at smaller scales. Recent research has demonstrated a connection between critical balance and dynamic alignment with intermittency \citep{Chandran_2015, Mallet_2017}. A more comprehensive analysis comparing the anisotropic scaling of higher-order moments to existing theories is in progress.\par 


When analyzing turbulence in the inner heliosphere, where the Alfv\'en speed approaches and sometimes exceed the solar wind speed special care must be used in applying homogeneous turbulence theories and models to the observed characteristics. This is especially important for power anisotropies, as in addition to wave-number couplings, the couplings to large scale gradients both in the radial and transverse directions may be fundamental, with the solar corona, for example, acting to refract energy in fast mode polarization into regions of low Alfv\'en speed or even providing some total reflection. These couplings could also affect spectral slopes in the parallel and perpendicular directions in the nascent solar wind \citep{Velli_91_waves}.

In conclusion, it is important to recognize the potential limitations of the current analysis, including the limited number of extended fast wind streams sampled by PSP and SO. These limitations may affect the statistical significance of the results and make it difficult to accurately determine the anisotropic scaling laws for these streams at lower frequencies. Therefore, it is advisable to continue collecting more samples from PSP and SO, particularly those of longer duration, to confirm the statistical significance of the findings. In addition, a more robust statistical analysis with longer intervals of data from Ulysses and Helios will be conducted to accurately determine the scaling of the anisotropy and its dependence on the heliocentric distance, phase of the solar cycle, and heliographic latitude.

\begin{acknowledgments}

This research was funded in part by the FIELDS experiment
on the Parker Solar Probe spacecraft, designed and developed under NASA contract
NNN06AA01C; the NASA Parker Solar Probe Observatory Scientist
grant NNX15AF34G and the  HERMES DRIVE NASA Science Center grant No. 80NSSC20K0604. 
The instruments of PSP were designed and developed under NASA contract NNN06AA01C. MV
acknowledges the support of ISSI, Bern, via the Johannes Geiss fellowship.

\software{Python \citep{van1995python}, SciPy \citep{2020SciPy-NMeth}, Pandas \citep{mckinney2010data},  Matplotlib \citep{Hunter2007Matplotlib}, }

\end{acknowledgments}

\bibliography{main}{}

\begin{thebibliography}{}
\expandafter\ifx\csname natexlab\endcsname\relax\def\natexlab#1{#1}\fi
\providecommand{\url}[1]{\href{#1}{#1}}
\providecommand{\dodoi}[1]{doi:~\href{http://doi.org/#1}{\nolinkurl{#1}}}
\providecommand{\doeprint}[1]{\href{http://ascl.net/#1}{\nolinkurl{http://ascl.net/#1}}}
\providecommand{\doarXiv}[1]{\href{https://arxiv.org/abs/#1}{\nolinkurl{https://arxiv.org/abs/#1}}}

\bibitem[{{Acu{\~n}a} {et~al.}(2008){Acu{\~n}a}, {Curtis}, {Scheifele},
  {Russell}, {Schroeder}, {Szabo}, \& {Luhmann}}]{Stereo_magnetometer}
{Acu{\~n}a}, M.~H., {Curtis}, D., {Scheifele}, J.~L., {et~al.} 2008, \ssr, 136,
  203, \dodoi{10.1007/s11214-007-9259-2}

\bibitem[{Adhikari {et~al.}(2022)Adhikari, Zank, Zhao, \&
  Telloni}]{Adhikari_2022}
Adhikari, L., Zank, G.~P., Zhao, L.-L., \& Telloni, D. 2022, The Astrophysical
  Journal, 933, 56, \dodoi{10.3847/1538-4357/ac70cb}

\bibitem[{{Alberti} {et~al.}(2022){Alberti}, {Benella}, {Consolini}, {Stumpo},
  \& {Benzi}}]{2022ApJ...940L..13A}
{Alberti}, T., {Benella}, S., {Consolini}, G., {Stumpo}, M., \& {Benzi}, R.
  2022, \apjl, 940, L13, \dodoi{10.3847/2041-8213/aca075}

\bibitem[{Alberti {et~al.}(2020)Alberti, Laurenza, Consolini, Milillo,
  Marcucci, Carbone, \& Bale}]{alberti_scaling_2020}
Alberti, T., Laurenza, M., Consolini, G., {et~al.} 2020, The Astrophysical
  Journal, 902, 84, \dodoi{10.3847/1538-4357/abb3d2}

\bibitem[{Bale {et~al.}(2016)Bale, Goetz, Harvey, Turin, Bonnell, Dudok~de Wit,
  Ergun, MacDowall, Pulupa, Andre, Bolton, Bougeret, Bowen, Burgess, Cattell,
  Chandran, Chaston, Chen, Choi, Connerney, Cranmer, Diaz-Aguado, Donakowski,
  Drake, Farrell, Fergeau, Fermin, Fischer, Fox, Glaser, Goldstein, Gordon,
  Hanson, Harris, Hayes, Hinze, Hollweg, Horbury, Howard, Hoxie, Jannet,
  Karlsson, Kasper, Kellogg, Kien, Klimchuk, Krasnoselskikh, Krucker, Lynch,
  Maksimovic, Malaspina, Marker, Martin, Martinez-Oliveros, McCauley, McComas,
  McDonald, Meyer-Vernet, Moncuquet, Monson, Mozer, Murphy, Odom, Oliverson,
  Olson, Parker, Pankow, Phan, Quataert, Quinn, Ruplin, Salem, Seitz, Sheppard,
  Siy, Stevens, Summers, Szabo, Timofeeva, Vaivads, Velli, Yehle, Werthimer, \&
  Wygant}]{bale_fields_2016}
Bale, S.~D., Goetz, K., Harvey, P.~R., {et~al.} 2016, ßr, 204, 49,
  \dodoi{10.1007/s11214-016-0244-5}

\bibitem[{{Bale} {et~al.}(2019){Bale}, {Badman}, {Bonnell}, {Bowen}, {Burgess},
  {Case}, {Cattell}, {Chandran}, {Chaston}, {Chen}, {Drake}, {de Wit},
  {Eastwood}, {Ergun}, {Farrell}, {Fong}, {Goetz}, {Goldstein}, {Goodrich},
  {Harvey}, {Horbury}, {Howes}, {Kasper}, {Kellogg}, {Klimchuk}, {Korreck},
  {Krasnoselskikh}, {Krucker}, {Laker}, {Larson}, {MacDowall}, {Maksimovic},
  {Malaspina}, {Martinez-Oliveros}, {McComas}, {Meyer-Vernet}, {Moncuquet},
  {Mozer}, {Phan}, {Pulupa}, {Raouafi}, {Salem}, {Stansby}, {Stevens}, {Szabo},
  {Velli}, {Woolley}, \& {Wygant}}]{2019Natur.576..237B}
{Bale}, S.~D., {Badman}, S.~T., {Bonnell}, J.~W., {et~al.} 2019, \nat, 576,
  237, \dodoi{10.1038/s41586-019-1818-7}

\bibitem[{Bandyopadhyay \& McComas(2021)}]{Bandyopadhyay_2021}
Bandyopadhyay, R., \& McComas, D.~J. 2021, The Astrophysical Journal, 923, 193,
  \dodoi{10.3847/1538-4357/ac3486}

\bibitem[{Belcher \& Davis~Jr.(1971)}]{belcher_large-amplitude_1971}
Belcher, J.~W., \& Davis~Jr., L. 1971, Journal of Geophysical Research
  (1896-1977), 76, 3534, \dodoi{https://doi.org/10.1029/JA076i016p03534}

\bibitem[{Beresnyak \& Lazarian(2010)}]{Beresnyak_2010}
Beresnyak, A., \& Lazarian, A. 2010, The Astrophysical Journal, 722, L110,
  \dodoi{10.1088/2041-8205/722/1/l110}

\bibitem[{Bieber {et~al.}(1996)Bieber, Wanner, \&
  Matthaeus}]{Bieber_anisotropy}
Bieber, J.~W., Wanner, W., \& Matthaeus, W.~H. 1996, Journal of Geophysical
  Research: Space Physics, 101, 2511, \dodoi{https://doi.org/10.1029/95JA02588}

\bibitem[{Biskamp(2003)}]{biskamp_magnetohydrodynamic_2003}
Biskamp, D. 2003, Magnetohydrodynamic {Turbulence}

\bibitem[{Boldyrev(2006)}]{boldyrev_2006}
Boldyrev, S. 2006, Phys. Rev. Lett., 96, 115002,
  \dodoi{10.1103/PhysRevLett.96.115002}

\bibitem[{Borovsky {et~al.}(2019)Borovsky, Denton, \& Smith}]{Borovsky_2012}
Borovsky, J.~E., Denton, M.~H., \& Smith, C.~W. 2019, Journal of Geophysical
  Research: Space Physics, 124, 2406,
  \dodoi{https://doi.org/10.1029/2019JA026580}

\bibitem[{Bowen {et~al.}(2020)Bowen, Bale, Bonnell, Dudok~de Wit, Goetz,
  Goodrich, Gruesbeck, Harvey, Jannet, Koval, MacDowall, Malaspina, Pulupa,
  Revillet, Sheppard, \& Szabo}]{https://doi.org/10.1029/2020JA027813}
Bowen, T.~A., Bale, S.~D., Bonnell, J.~W., {et~al.} 2020, Journal of
  Geophysical Research: Space Physics, 125, e2020JA027813,
  \dodoi{https://doi.org/10.1029/2020JA027813}

\bibitem[{Bruno {et~al.}(2003)Bruno, Carbone, Sorriso-Valvo, \&
  Bavassano}]{bruno_radial_2003}
Bruno, R., Carbone, V., Sorriso-Valvo, L., \& Bavassano, B. 2003, Journal of
  Geophysical Research (Space Physics), 108, 1130, \dodoi{10.1029/2002JA009615}

\bibitem[{Chandran(2018)}]{chandran_2018}
Chandran, B. D.~G. 2018, Journal of Plasma Physics, 84, 905840106,
  \dodoi{10.1017/S0022377818000016}

\bibitem[{{Chandran} \& {Perez}(2019)}]{Chandran_Perez_2019}
{Chandran}, B. D.~G., \& {Perez}, J.~C. 2019, Journal of Plasma Physics, 85,
  905850409, \dodoi{10.1017/S0022377819000540}

\bibitem[{{Chandran} {et~al.}(2015){Chandran}, {Schekochihin}, \&
  {Mallet}}]{Chandran_2015}
{Chandran}, B.~D.~G., {Schekochihin}, A.~A., \& {Mallet}, A. 2015, \apj, 807,
  39, \dodoi{10.1088/0004-637X/807/1/39}

\bibitem[{Chen {et~al.}(2010)Chen, Horbury, Schekochihin, Wicks, Alexandrova,
  \& Mitchell}]{Chen_2010_anisotropy_multispacecraft}
Chen, C. H.~K., Horbury, T.~S., Schekochihin, A.~A., {et~al.} 2010, Phys. Rev.
  Lett., 104, 255002, \dodoi{10.1103/PhysRevLett.104.255002}

\bibitem[{{Chen} {et~al.}(2014){Chen}, {Leung}, {Boldyrev}, {Maruca}, \&
  {Bale}}]{2014GeoRL..41.8081C}
{Chen}, C.~H.~K., {Leung}, L., {Boldyrev}, S., {Maruca}, B.~A., \& {Bale},
  S.~D. 2014, \grl, 41, 8081, \dodoi{10.1002/2014GL062009}

\bibitem[{{Chen} {et~al.}(2011){Chen}, {Mallet}, {Yousef}, {Schekochihin}, \&
  {Horbury}}]{Chen-anisotropic_2011}
{Chen}, C.~H.~K., {Mallet}, A., {Yousef}, T.~A., {Schekochihin}, A.~A., \&
  {Horbury}, T.~S. 2011, \mnras, 415, 3219,
  \dodoi{10.1111/j.1365-2966.2011.18933.x}

\bibitem[{Chen {et~al.}(2020)Chen, Bale, Bonnell, Borovikov, Bowen, Burgess,
  Case, Chandran, Wit, Goetz, Harvey, Kasper, Klein, Korreck, Larson, Livi,
  MacDowall, Malaspina, Mallet, McManus, Moncuquet, Pulupa, Stevens, \&
  Whittlesey}]{chen_evolution_2020}
Chen, C. H.~K., Bale, S.~D., Bonnell, J.~W., {et~al.} 2020, The Astrophysical
  Journal Supplement Series, 246, 53, \dodoi{10.3847/1538-4365/ab60a3}

\bibitem[{Chhiber(2022)}]{Chhiber_2022}
Chhiber, R. 2022, The Astrophysical Journal, 939, 33,
  \dodoi{10.3847/1538-4357/ac9386}

\bibitem[{{Cho} \& {Vishniac}(2000)}]{Cho_Vishniac_2000}
{Cho}, J., \& {Vishniac}, E.~T. 2000, \apj, 539, 273, \dodoi{10.1086/309213}

\bibitem[{{Cuesta} {et~al.}(2022){Cuesta}, {Chhiber}, {Roy}, {Goodwill},
  {Pecora}, {Jarosik}, {Matthaeus}, {Parashar}, \&
  {Bandyopadhyay}}]{Cuesta_anisotropy}
{Cuesta}, M.~E., {Chhiber}, R., {Roy}, S., {et~al.} 2022, \apjl, 932, L11,
  \dodoi{10.3847/2041-8213/ac73fd}

\bibitem[{{Dasso} {et~al.}(2005){Dasso}, {Milano}, {Matthaeus}, \&
  {Smith}}]{Dasso_2005}
{Dasso}, S., {Milano}, L.~J., {Matthaeus}, W.~H., \& {Smith}, C.~W. 2005,
  \apjl, 635, L181, \dodoi{10.1086/499559}

\bibitem[{Davies \& Gather(1993)}]{davies_identification_1993}
Davies, L., \& Gather, U. 1993, Journal of the American Statistical
  Association, 88, 782, \dodoi{10.1080/01621459.1993.10476339}

\bibitem[{Davis {et~al.}(2023)Davis, Chandran, Bowen, Badman, de~Wit, Chen,
  Bale, Huang, Sioulas, \& Velli}]{davis2023}
Davis, N., Chandran, B. D.~G., Bowen, T.~A., {et~al.} 2023, The Evolution of
  the 1/f Range Within a Single Fast-Solar-Wind Stream Between 17.4 and 45.7
  Solar Radii.
\newblock \doarXiv{2303.01663}

\bibitem[{Dong {et~al.}(2022)Dong, Wang, Huang, Comisso, Sandstrom, \&
  Bhattacharjee}]{Dong_22_largest_mhd_turb}
Dong, C., Wang, L., Huang, Y.-M., {et~al.} 2022, Science Advances, 8, eabn7627,
  \dodoi{10.1126/sciadv.abn7627}

\bibitem[{Duan {et~al.}(2021)Duan, He, Bowen, Woodham, Wang, Chen, Mallet, \&
  Bale}]{Duan_2021}
Duan, D., He, J., Bowen, T.~A., {et~al.} 2021, The Astrophysical Journal
  Letters, 915, L8, \dodoi{10.3847/2041-8213/ac07ac}

\bibitem[{{Dudok de Wit} {et~al.}(2013){Dudok de Wit}, {Alexandrova}, {Furno},
  {Sorriso-Valvo}, \& {Zimbardo}}]{Dudok_de_wit_Samples_Rule}
{Dudok de Wit}, T., {Alexandrova}, O., {Furno}, I., {Sorriso-Valvo}, L., \&
  {Zimbardo}, G. 2013, \ssr, 178, 665, \dodoi{10.1007/s11214-013-9974-9}

\bibitem[{Elsasser(1950)}]{elsasser_1950}
Elsasser, W.~M. 1950, Phys. Rev., 79, 183, \dodoi{10.1103/PhysRev.79.183}

\bibitem[{{Galeev} \& {Oraevskii}(1963)}]{Galeev_1963_PDI}
{Galeev}, A.~A., \& {Oraevskii}, V.~N. 1963, Soviet Physics Doklady, 7, 988

\bibitem[{{Galtier} {et~al.}(2000){Galtier}, {Nazarenko}, {Newell}, \&
  {Pouquet}}]{Galtier_2000_anisotropic}
{Galtier}, S., {Nazarenko}, S.~V., {Newell}, A.~C., \& {Pouquet}, A. 2000,
  Journal of Plasma Physics, 63, 447, \dodoi{10.1017/S0022377899008284}

\bibitem[{Gerick {et~al.}(2017)Gerick, Saur, \& von Papen}]{Gerick_2017}
Gerick, F., Saur, J., \& von Papen, M. 2017, The Astrophysical Journal, 843, 5,
  \dodoi{10.3847/1538-4357/aa767c}

\bibitem[{Goldreich \& Sridhar(1995)}]{goldreich_toward_1995}
Goldreich, P., \& Sridhar, S. 1995, {\textbackslash}apj, 438, 763,
  \dodoi{10.1086/175121}

\bibitem[{Goldreich \& Sridhar(1997)}]{goldreich_magnetohydrodynamic_1997}
---. 1997, {\textbackslash}apj, 485, 680, \dodoi{10.1086/304442}

\bibitem[{Gurland \& Tripathi(1971)}]{10.2307/2682923}
Gurland, J., \& Tripathi, R.~C. 1971, The American Statistician, 25, 30.
\newblock \url{http://www.jstor.org/stable/2682923}

\bibitem[{He {et~al.}(2013)He, Tu, Marsch, Bourouaine, \& Pei}]{He_2013}
He, J., Tu, C., Marsch, E., Bourouaine, S., \& Pei, Z. 2013, The Astrophysical
  Journal, 773, 72, \dodoi{10.1088/0004-637X/773/1/72}

\bibitem[{{Higdon}(1984)}]{higdon_anisotropic}
{Higdon}, J.~C. 1984, \apj, 285, 109, \dodoi{10.1086/162481}

\bibitem[{Horbury {et~al.}(2008)Horbury, Forman, \&
  Oughton}]{horbury_anisotropic_2008}
Horbury, T.~S., Forman, M., \& Oughton, S. 2008, {\textbackslash}prl, 101,
  175005, \dodoi{10.1103/PhysRevLett.101.175005}

\bibitem[{{Horbury} {et~al.}(2012){Horbury}, {Wicks}, \&
  {Chen}}]{Horbury_anisotropy}
{Horbury}, T.~S., {Wicks}, R.~T., \& {Chen}, C.~H.~K. 2012,
  {\textbackslash}ssr, 172, 325, \dodoi{10.1007/s11214-011-9821-9}

\bibitem[{Horbury {et~al.}(2020)Horbury, O'Brien, Carrasco~Blazquez, Bendyk,
  Brown, Hudson, Evans, Oddy, Carr, Beek, Cupido, Bhattacharya, Dominguez,
  Matthews, Myklebust, Whiteside, Bale, Baumjohann, Burgess, Carbone, Cargill,
  Eastwood, Erdös, Fletcher, Forsyth, Giacalone, Glassmeier, Goldstein,
  Hoeksema, Lockwood, Magnes, Maksimovic, Marsch, Matthaeus, Murphy,
  Nakariakov, Owen, Owens, Rodriguez-Pacheco, Richter, Riley, Russell,
  Schwartz, Vainio, Velli, Vennerstrom, Walsh, Wimmer-Schweingruber, Zank,
  Müller, Zouganelis, \& Walsh}]{horbury_solar_2020}
Horbury, T.~S., O'Brien, H., Carrasco~Blazquez, I., {et~al.} 2020, åp, 642,
  A9, \dodoi{10.1051/0004-6361/201937257}

\bibitem[{Huang {et~al.}(2022)Huang, Xu, Zhang, Sahraoui, Andrés, He, Yuan,
  Deng, Jiang, Wei, Xiong, Wang, Yu, \& Lin}]{Huang_2022}
Huang, S.~Y., Xu, S.~B., Zhang, J., {et~al.} 2022, The Astrophysical Journal
  Letters, 929, L6, \dodoi{10.3847/2041-8213/ac5f02}

\bibitem[{Huang {et~al.}(2023)Huang, Sioulas, Shi, Velli, Bowen, Davis,
  Chandran, Kang, Shi, Huang, Bale, Kasper, Larson, Livi, Whittlesey, Rahmati,
  Paulson, Stevens, Case, de~Wit, Malaspina, Bonnell, Goetz, Harvey, \&
  MacDowall}]{huang2023}
Huang, Z., Sioulas, N., Shi, C., {et~al.} 2023, New Observations of Solar Wind
  1/f Turbulence Spectrum from Parker Solar Probe.
\newblock \doarXiv{2303.00843}

\bibitem[{Hunter(2007)}]{Hunter2007Matplotlib}
Hunter, J.~D. 2007, Computing in Science \& Engineering, 9, 90,
  \dodoi{10.1109/MCSE.2007.55}

\bibitem[{Iroshnikov(1963)}]{iroshnikov_turbulence_1963}
Iroshnikov, P.~S. 1963, Astronomicheskii Zhurnal, 40, 742.
\newblock \url{https://ui.adsabs.harvard.edu/abs/1963AZh....40..742I}

\bibitem[{Kasper {et~al.}(2016)Kasper, Abiad, Austin, Balat-Pichelin, Bale,
  Belcher, Berg, Bergner, Berthomier, Bookbinder, Brodu, Caldwell, Case,
  Chandran, Cheimets, Cirtain, Cranmer, Curtis, Daigneau, Dalton, Dasgupta,
  DeTomaso, Diaz-Aguado, Djordjevic, Donaskowski, Effinger, Florinski, Fox,
  Freeman, Gallagher, Gary, Gauron, Gates, Goldstein, Golub, Gordon, Gurnee,
  Guth, Halekas, Hatch, Heerikuisen, Ho, Hu, Johnson, Jordan, Korreck, Larson,
  Lazarus, Li, Livi, Ludlam, Maksimovic, McFadden, Marchant, Maruca, McComas,
  Messina, Mercer, Park, Peddie, Pogorelov, Reinhart, Richardson, Robinson,
  Rosen, Skoug, Slagle, Steinberg, Stevens, Szabo, Taylor, Tiu, Turin, Velli,
  Webb, Whittlesey, Wright, Wu, \& Zank}]{kasper_solar_2016}
Kasper, J.~C., Abiad, R., Austin, G., {et~al.} 2016, ßr, 204, 131,
  \dodoi{10.1007/s11214-015-0206-3}

\bibitem[{Klein {et~al.}(2015)Klein, Perez, Verscharen, Mallet, \&
  Chandran}]{Klein_2015}
Klein, K.~G., Perez, J.~C., Verscharen, D., Mallet, A., \& Chandran, B. D.~G.
  2015, The Astrophysical Journal, 801, L18,
  \dodoi{10.1088/2041-8205/801/1/l18}

\bibitem[{Kraichnan(1965)}]{kraichnan_inertial-range_1965}
Kraichnan, R.~H. 1965, The Physics of Fluids, 8, 1385

\bibitem[{{Lithwick} {et~al.}(2007){Lithwick}, {Goldreich}, \&
  {Sridhar}}]{Lithwick_2007_imbalanced_critical_balance}
{Lithwick}, Y., {Goldreich}, P., \& {Sridhar}, S. 2007, \apj, 655, 269,
  \dodoi{10.1086/509884}

\bibitem[{{Maksimovic} {et~al.}(2020){Maksimovic}, {Bale},
  {Ber{\v{c}}i{\v{c}}}, {Bonnell}, {Case}, {Dudok de Wit}, {Goetz}, {Halekas},
  {Harvey}, {Issautier}, {Kasper}, {Korreck}, {Jagarlamudi}, {Lahmiti},
  {Larson}, {Lecacheux}, {Livi}, {MacDowall}, {Malaspina}, {Martinovi{\'c}},
  {Meyer-Vernet}, {Moncuquet}, {Pulupa}, {Salem}, {Stevens},
  {{\v{S}}tver{\'a}k}, {Velli}, \& {Whittlesey}}]{2020ApJS..246...62M}
{Maksimovic}, M., {Bale}, S.~D., {Ber{\v{c}}i{\v{c}}}, L., {et~al.} 2020,
  \apjs, 246, 62, \dodoi{10.3847/1538-4365/ab61fc}

\bibitem[{Malara {et~al.}(2022)Malara, Primavera, \& Veltri}]{Malara_22_PDI}
Malara, F., Primavera, L., \& Veltri, P. 2022, Universe, 8,
  \dodoi{10.3390/universe8080391}

\bibitem[{{Mallet} \& {Schekochihin}(2017)}]{Mallet_2017}
{Mallet}, A., \& {Schekochihin}, A.~A. 2017, \mnras, 466, 3918,
  \dodoi{10.1093/mnras/stw3251}

\bibitem[{{Maron} \& {Goldreich}(2001)}]{Maron_2001}
{Maron}, J., \& {Goldreich}, P. 2001, \apj, 554, 1175, \dodoi{10.1086/321413}

\bibitem[{Mason {et~al.}(2006)Mason, Cattaneo, \& Boldyrev}]{masson_2006}
Mason, J., Cattaneo, F., \& Boldyrev, S. 2006, Phys. Rev. Lett., 97, 255002,
  \dodoi{10.1103/PhysRevLett.97.255002}

\bibitem[{Matteini {et~al.}(2014)Matteini, Horbury, Neugebauer, \&
  Goldstein}]{Matteini_2014}
Matteini, L., Horbury, T.~S., Neugebauer, M., \& Goldstein, B.~E. 2014,
  Geophysical Research Letters, 41, 259,
  \dodoi{https://doi.org/10.1002/2013GL058482}

\bibitem[{{Matthaeus} {et~al.}(1990){Matthaeus}, {Goldstein}, \&
  {Roberts}}]{Matthaeus_1990_anisotropy}
{Matthaeus}, W.~H., {Goldstein}, M.~L., \& {Roberts}, D.~A. 1990, \jgr, 95,
  20673, \dodoi{10.1029/JA095iA12p20673}

\bibitem[{McKinney {et~al.}(2010)}]{mckinney2010data}
McKinney, W., {et~al.} 2010, in Proceedings of the 9th Python in Science
  Conference, Vol. 445, Austin, TX, 51--56

\bibitem[{Meyrand {et~al.}(2016)Meyrand, Galtier, \&
  Kiyani}]{Meyrand_transition}
Meyrand, R., Galtier, S., \& Kiyani, K.~H. 2016, Phys. Rev. Lett., 116, 105002,
  \dodoi{10.1103/PhysRevLett.116.105002}

\bibitem[{Moncuquet {et~al.}(2020)Moncuquet, Meyer-Vernet, Issautier, Pulupa,
  Bonnell, Bale, de~Wit, Goetz, Griton, Harvey, MacDowall, Maksimovic, \&
  Malaspina}]{Moncuquet_2020}
Moncuquet, M., Meyer-Vernet, N., Issautier, K., {et~al.} 2020, The
  Astrophysical Journal Supplement Series, 246, 44,
  \dodoi{10.3847/1538-4365/ab5a84}

\bibitem[{{Montgomery} \& {Matthaeus}(1995)}]{1995ApJ...447..706M}
{Montgomery}, D., \& {Matthaeus}, W.~H. 1995, \apj, 447, 706,
  \dodoi{10.1086/175910}

\bibitem[{Montgomery \& Turner(1981)}]{Montgomery_turner_1981}
Montgomery, D., \& Turner, L. 1981, The Physics of Fluids, 24, 825,
  \dodoi{10.1063/1.863455}

\bibitem[{{Ng} \& {Bhattacharjee}(1996)}]{Ng_Bhattacharjee}
{Ng}, C.~S., \& {Bhattacharjee}, A. 1996, \apj, 465, 845,
  \dodoi{10.1086/177468}

\bibitem[{Němeček {et~al.}(2021)Němeček, Šafránková, Němec,
  Ďurovcová, Pitňa, Alterman, Voitenko, Pavlů, \& Stevens}]{Nemecek_2021}
Němeček, Z., Šafránková, J., Němec, F., {et~al.} 2021, Atmosphere, 12,
  1277, \dodoi{10.3390/atmos12101277}

\bibitem[{{Oboukhov}(1962)}]{oboukhov_1962}
{Oboukhov}, A.~M. 1962, Journal of Fluid Mechanics, 13, 77,
  \dodoi{10.1017/S0022112062000506}

\bibitem[{Osman {et~al.}(2012)Osman, Matthaeus, Wan, \&
  Rappazzo}]{osman_intermittency_2012}
Osman, K.~T., Matthaeus, W.~H., Wan, M., \& Rappazzo, A.~F. 2012, Phys. Rev.
  Lett., 108, 261102, \dodoi{10.1103/PhysRevLett.108.261102}

\bibitem[{Oughton {et~al.}(2015)Oughton, Matthaeus, Wan, \&
  Osman}]{Oughton_anisotropy_review}
Oughton, S., Matthaeus, W.~H., Wan, M., \& Osman, K.~T. 2015, Philosophical
  Transactions of the Royal Society A: Mathematical, Physical and Engineering
  Sciences, 373, 20140152, \dodoi{10.1098/rsta.2014.0152}

\bibitem[{Owen {et~al.}(2020)Owen, Bruno, Livi, Louarn, Al~Janabi, Allegrini,
  Amoros, Baruah, Barthe, Berthomier, Bordon, Brockley-Blatt, Brysbaert,
  Capuano, Collier, DeMarco, Fedorov, Ford, Fortunato, Fratter, Galvin,
  Hancock, Heirtzler, Kataria, Kistler, Lepri, Lewis, Loeffler, Marty, Mathon,
  Mayall, Mele, Ogasawara, Orlandi, Pacros, Penou, Persyn, Petiot, Phillips,
  Přech, Raines, Reden, Rouillard, Rousseau, Rubiella, Seran, Spencer, Thomas,
  Trevino, Verscharen, Wurz, Alapide, Amoruso, André, Anekallu, Arciuli,
  Arnett, Ascolese, Bancroft, Bland, Brysch, Calvanese, Castronuovo, Čermák,
  Chornay, Clemens, Coker, Collinson, D'Amicis, Dandouras, Darnley, Davies,
  Davison, De~Los~Santos, Devoto, Dirks, Edlund, Fazakerley, Ferris, Frost,
  Fruit, Garat, Génot, Gibson, Gilbert, de~Giosa, Gradone, Hailey, Horbury,
  Hunt, Jacquey, Johnson, Lavraud, Lawrenson, Leblanc, Lockhart, Maksimovic,
  Malpus, Marcucci, Mazelle, Monti, Myers, Nguyen, Rodriguez-Pacheco, Phillips,
  Popecki, Rees, Rogacki, Ruane, Rust, Salatti, Sauvaud, Stakhiv, Stange,
  Stubbs, Taylor, Techer, Terrier, Thibodeaux, Urdiales, Varsani, Walsh,
  Watson, Wheeler, Willis, Wimmer-Schweingruber, Winter, Yardley, \&
  Zouganelis}]{owen_solar_2020}
Owen, C.~J., Bruno, R., Livi, S., {et~al.} 2020, åp, 642, A16,
  \dodoi{10.1051/0004-6361/201937259}

\bibitem[{{Parker}(1979)}]{parker_magnetic_fields}
{Parker}, E.~N. 1979, {Cosmical magnetic fields. Their origin and their
  activity}

\bibitem[{Perez \& Boldyrev(2009)}]{Perez_Boldyrev_Extend}
Perez, J.~C., \& Boldyrev, S. 2009, Phys. Rev. Lett., 102, 025003,
  \dodoi{10.1103/PhysRevLett.102.025003}

\bibitem[{{Perez} {et~al.}(2021){Perez}, {Bourouaine}, {Chen}, \&
  {Raouafi}}]{2021A&A...650A..22P}
{Perez}, J.~C., {Bourouaine}, S., {Chen}, C. H.~K., \& {Raouafi}, N.~E. 2021,
  \aap, 650, A22, \dodoi{10.1051/0004-6361/202039879}

\bibitem[{{Pi} {et~al.}(2020){Pi}, {Pit{\r{A}}a}, {N{\v{e}}me{\v{c}}ek},
  {{\v{S}}afr{\'a}nkov{\'a}}, {Shue}, \& {Yang}}]{Pi_2020}
{Pi}, G., {Pit{\r{A}}a}, A., {N{\v{e}}me{\v{c}}ek}, Z., {et~al.} 2020,
  \solphys, 295, 84, \dodoi{10.1007/s11207-020-01646-8}

\bibitem[{{Pine} {et~al.}(2020){Pine}, {Smith}, {Hollick}, {Argall}, {Vasquez},
  {Isenberg}, {Schwadron}, {Joyce}, {Sok{\'o}{\l}}, {Bzowski}, {Kubiak},
  {Hamilton}, {McLaurin}, \& {Leamon}}]{Pine_2020}
{Pine}, Z.~B., {Smith}, C.~W., {Hollick}, S.~J., {et~al.} 2020, \apj, 900, 93,
  \dodoi{10.3847/1538-4357/abab11}

\bibitem[{Podesta(2009)}]{Podesta_2009}
Podesta, J.~J. 2009, The Astrophysical Journal, 698, 986,
  \dodoi{10.1088/0004-637x/698/2/986}

\bibitem[{{Schekochihin}(2022)}]{Schekochihin_2022}
{Schekochihin}, A.~A. 2022, Journal of Plasma Physics, 88, 155880501,
  \dodoi{10.1017/S0022377822000721}

\bibitem[{Schekochihin {et~al.}(2009)Schekochihin, Cowley, Dorland, Hammett,
  Howes, Quataert, \& Tatsuno}]{Schekochihin_2009_review}
Schekochihin, A.~A., Cowley, S.~C., Dorland, W., {et~al.} 2009, The
  Astrophysical Journal Supplement Series, 182, 310,
  \dodoi{10.1088/0067-0049/182/1/310}

\bibitem[{Shebalin {et~al.}(1983)Shebalin, Matthaeus, \&
  Montgomery}]{shebalin_matthaeus_montgomery_1983}
Shebalin, J.~V., Matthaeus, W.~H., \& Montgomery, D. 1983, Journal of Plasma
  Physics, 29, 525–547, \dodoi{10.1017/S0022377800000933}

\bibitem[{Shi {et~al.}(2021)Shi, Velli, Panasenco, Tenerani, Réville, Bale,
  Kasper, Korreck, Bonnell, Dudok~de Wit, Malaspina, Goetz, Harvey, MacDowall,
  Pulupa, Case, Larson, Verniero, Livi, Stevens, Whittlesey, Maksimovic, \&
  Moncuquet}]{shi_alfvenic_2021}
Shi, C., Velli, M., Panasenco, O., {et~al.} 2021, åp, 650, A21,
  \dodoi{10.1051/0004-6361/202039818}

\bibitem[{{Shi} {et~al.}(2023){Shi}, {Velli}, {Lionello}, {Sioulas}, {Huang},
  {Halekas}, {Tenerani}, {R{\'e}ville}, {Dakeyo}, {Maksimovi{\'c}}, \&
  {Bale}}]{2023_Shi}
{Shi}, C., {Velli}, M., {Lionello}, R., {et~al.} 2023, arXiv e-prints,
  arXiv:2301.00852.
\newblock \doarXiv{2301.00852}

\bibitem[{Sioulas {et~al.}(2022{\natexlab{a}})Sioulas, Huang, Velli, Chhiber,
  Cuesta, Shi, Matthaeus, Bandyopadhyay, Vlahos, Bowen, Qudsi, Bale, Owen,
  Louarn, Fedorov, Maksimovi{\'{c}}, Stevens, Case, Kasper, Larson, Pulupa, \&
  Livi}]{Sioulas_2022_intermittency}
Sioulas, N., Huang, Z., Velli, M., {et~al.} 2022{\natexlab{a}}, The
  Astrophysical Journal, 934, 143, \dodoi{10.3847/1538-4357/ac7aa2}

\bibitem[{Sioulas {et~al.}(2022{\natexlab{b}})Sioulas, Huang, Shi, Velli,
  Tenerani, Vlahos, Bowen, Bale, Bonnell, Harvey, Larson, Pulupa, Livi,
  Woodham, Horbury, Stevens, de~Wit, MacDowall, Malaspina, Goetz, Huang,
  Kasper, Owen, Maksimović, Louarn, \& Fedorov}]{sioulas_turb_22_no1}
Sioulas, N., Huang, Z., Shi, C., {et~al.} 2022{\natexlab{b}}, Magnetic field
  spectral evolution in the inner heliosphere,  arXiv,
  \dodoi{10.48550/ARXIV.2209.02451}

\bibitem[{Sioulas {et~al.}(2022{\natexlab{c}})Sioulas, Huang, Shi, Velli,
  Tenerani, Vlahos, Bowen, Bale, Bonnell, Harvey, Larson, Pulupa, Livi,
  Woodham, Horbury, Stevens, de~Wit, MacDowall, Malaspina, Goetz, Huang,
  Kasper, Owen, Maksimović, Louarn, \&
  Fedorov}]{Sioulas_2022_spectral_evolution}
---. 2022{\natexlab{c}}, Magnetic field spectral evolution in the inner
  heliosphere,  arXiv, \dodoi{10.48550/ARXIV.2209.02451}

\bibitem[{{Sridhar} \& {Goldreich}(1994)}]{1994_goldreich}
{Sridhar}, S., \& {Goldreich}, P. 1994, \apj, 432, 612, \dodoi{10.1086/174600}

\bibitem[{Taylor(1938)}]{taylor_spectrum_1938}
Taylor, G.~I. 1938, Proceedings of the Royal Society of London. Series A -
  Mathematical and Physical Sciences, 164, 476, \dodoi{10.1098/rspa.1938.0032}

\bibitem[{Telloni(2022)}]{telloni_transition}
Telloni, D. 2022, Frontiers in Astronomy and Space Sciences, 9,
  \dodoi{10.3389/fspas.2022.917393}

\bibitem[{Telloni {et~al.}(2021)Telloni, Sorriso-Valvo, Woodham, Panasenco,
  Velli, Carbone, Zank, Bruno, Perrone, Nakanotani, Shi, D'Amicis, De~Marco,
  Jagarlamudi, Steinvall, Marino, Adhikari, Zhao, Liang, Tenerani, Laker,
  Horbury, Bale, Pulupa, Malaspina, MacDowall, Goetz, de~Wit, Harvey, Kasper,
  Korreck, Larson, Case, Stevens, Whittlesey, Livi, Owen, Livi, Louarn,
  Antonucci, Romoli, O'Brien, Evans, \& Angelini}]{telloni_evolution_2021}
Telloni, D., Sorriso-Valvo, L., Woodham, L.~D., {et~al.} 2021, \apjl, 912, L21,
  \dodoi{10.3847/2041-8213/abf7d1}

\bibitem[{{Tenerani} {et~al.}(2017){Tenerani}, {Velli}, \&
  {Hellinger}}]{Tenerani_2017ApJ_PDI}
{Tenerani}, A., {Velli}, M., \& {Hellinger}, P. 2017, \apj, 851, 99,
  \dodoi{10.3847/1538-4357/aa9bef}

\bibitem[{Van~Rossum \& Drake~Jr(1995)}]{van1995python}
Van~Rossum, G., \& Drake~Jr, F.~L. 1995, Python reference manual (Centrum voor
  Wiskunde en Informatica Amsterdam)

\bibitem[{Vasquez {et~al.}(2007)Vasquez, Smith, Hamilton, MacBride, \&
  Leamon}]{Vasquez_2007}
Vasquez, B.~J., Smith, C.~W., Hamilton, K., MacBride, B.~T., \& Leamon, R.~J.
  2007, Journal of Geophysical Research: Space Physics, 112,
  \dodoi{https://doi.org/10.1029/2007JA012305}

\bibitem[{{Vech} {et~al.}(2018){Vech}, {Mallet}, {Klein}, \&
  {Kasper}}]{2018_vech}
{Vech}, D., {Mallet}, A., {Klein}, K.~G., \& {Kasper}, J.~C. 2018, \apjl, 855,
  L27, \dodoi{10.3847/2041-8213/aab351}

\bibitem[{{Velli}(1993)}]{Velli_93}
{Velli}, M. 1993, A\&A, 270, 304

\bibitem[{Velli {et~al.}(1991)Velli, Grappin, \& Mangeney}]{Velli_91_waves}
Velli, M., Grappin, R., \& Mangeney, A. 1991, Geophysical \& Astrophysical
  Fluid Dynamics, 62, 101, \dodoi{10.1080/03091929108229128}

\bibitem[{{Verdini} {et~al.}(2018){Verdini}, {Grappin}, {Alexandrova}, \&
  {Lion}}]{Verdini_2018}
{Verdini}, A., {Grappin}, R., {Alexandrova}, O., \& {Lion}, S. 2018, \apj, 853,
  85, \dodoi{10.3847/1538-4357/aaa433}

\bibitem[{Virtanen {et~al.}(2020)Virtanen, Gommers, Oliphant, Haberland, Reddy,
  Cournapeau, Burovski, Peterson, Weckesser, Bright, {van der Walt}, Brett,
  Wilson, Millman, Mayorov, Nelson, Jones, Kern, Larson, Carey, Polat, Feng,
  Moore, {VanderPlas}, Laxalde, Perktold, Cimrman, Henriksen, Quintero, Harris,
  Archibald, Ribeiro, Pedregosa, {van Mulbregt}, \& {SciPy 1.0
  Contributors}}]{2020SciPy-NMeth}
Virtanen, P., Gommers, R., Oliphant, T.~E., {et~al.} 2020, Nature Methods, 17,
  261, \dodoi{10.1038/s41592-019-0686-2}

\bibitem[{{Wang} {et~al.}(2020){Wang}, {He}, {Alexandrova}, {Dunlop}, \&
  {Perrone}}]{WANG_2020}
{Wang}, T., {He}, J., {Alexandrova}, O., {Dunlop}, M., \& {Perrone}, D. 2020,
  \apj, 898, 91, \dodoi{10.3847/1538-4357/ab99ca}

\bibitem[{Weygand {et~al.}(2009)Weygand, Matthaeus, Dasso, Kivelson, Kistler,
  \& Mouikis}]{Weygand_anisotropy}
Weygand, J.~M., Matthaeus, W.~H., Dasso, S., {et~al.} 2009, Journal of
  Geophysical Research: Space Physics, 114,
  \dodoi{https://doi.org/10.1029/2008JA013766}

\bibitem[{{Wicks} {et~al.}(2010){Wicks}, {Horbury}, {Chen}, \&
  {Schekochihin}}]{wicks_anisotropy1}
{Wicks}, R.~T., {Horbury}, T.~S., {Chen}, C.~H.~K., \& {Schekochihin}, A.~A.
  2010, mnras, 407, L31, \dodoi{10.1111/j.1745-3933.2010.00898.x}

\bibitem[{Wicks {et~al.}(2013)Wicks, Mallet, Horbury, Chen, Schekochihin, \&
  Mitchell}]{wicks_alignmene_2013}
Wicks, R.~T., Mallet, A., Horbury, T.~S., {et~al.} 2013, Phys. Rev. Lett., 110,
  025003, \dodoi{10.1103/PhysRevLett.110.025003}

\bibitem[{Woodham(2019)}]{phdthesis}
Woodham, L. 2019, PhD thesis, \dodoi{10.13140/RG.2.2.21508.45443}

\bibitem[{Wu {et~al.}(2022)Wu, He, Yang, Wang, Huang, \& Yuan}]{Wu2022OnTS}
Wu, H., He, J., Yang, L., {et~al.} 2022, On the scaling and anisotropy of two
  subranges in the inertial range of solar wind turbulence,  arXiv,
  \dodoi{10.48550/ARXIV.2209.12409}

\bibitem[{Zank {et~al.}(2020)Zank, Nakanotani, Zhao, Adhikari, \&
  Telloni}]{Zank_2020_transition}
Zank, G.~P., Nakanotani, M., Zhao, L.-L., Adhikari, L., \& Telloni, D. 2020,
  The Astrophysical Journal, 900, 115, \dodoi{10.3847/1538-4357/abad30}

\bibitem[{{Zank} {et~al.}(2022{\natexlab{a}}){Zank}, {Zhao}, {Adhikari},
  {Telloni}, {Kasper}, {Stevens}, {Rahmati}, \& {Bale}}]{Zank_2022}
{Zank}, G.~P., {Zhao}, L.~L., {Adhikari}, L., {et~al.} 2022{\natexlab{a}},
  \apjl, 926, L16, \dodoi{10.3847/2041-8213/ac51da}

\bibitem[{{Zank} {et~al.}(2022{\natexlab{b}}){Zank}, {Zhao}, {Adhikari},
  {Telloni}, {Kasper}, {Stevens}, {Rahmati}, \& {Bale}}]{2022_Zank}
---. 2022{\natexlab{b}}, \apjl, 926, L16, \dodoi{10.3847/2041-8213/ac51da}

\bibitem[{{Zhao} {et~al.}(2022){Zhao}, {Zank}, {Adhikari}, \&
  {Nakanotani}}]{Zhao_2022ApJ}
{Zhao}, L.~L., {Zank}, G.~P., {Adhikari}, L., \& {Nakanotani}, M. 2022, \apjl,
  924, L5, \dodoi{10.3847/2041-8213/ac4415}

\bibitem[{{Zhao} {et~al.}(2020){Zhao}, {Zank}, {Adhikari}, {Nakanotani},
  {Telloni}, \& {Carbone}}]{2020ApJ...898..113Z}
{Zhao}, L.~L., {Zank}, G.~P., {Adhikari}, L., {et~al.} 2020, \apj, 898, 113,
  \dodoi{10.3847/1538-4357/ab9b7e}

\end{thebibliography}
\bibliographystyle{aasjournal}

\end{CJK*}
\end{document}